\catcode`\@=11

\newif\if@fewtab\@fewtabtrue
{\count255=\time\divide\count255 by 60
\xdef\hourmin{\number\count255}
\multiply\count255 by-60\advance\count255 by\time
\xdef\hourmin{\hourmin:\ifnum\count255<10 0\fi\the\count255}}
\def\ps@draft{\let\@mkboth\@gobbletwo
    \def\@oddhead{}
    \def\@oddfoot{\hbox to 7 cm{\tiny \versionno
       \hfil}\hskip -7cm\hfil\rm\thepage \hfil {\tiny\draftdate}}
    \def\@evenhead{}\let\@evenfoot\@oddfoot}

\def\draftcite#1{\ifnum\draftcontrol=1#1\else{}\fi}
\def\@lbibitem[#1]#2{\item{}\hskip -3cm \hbox to 2cm
{\hfil$\scriptstyle\draftcite{#2}$}\hskip
1cm[\@biblabel{#1}]\if@filesw
     {\def\protect##1{\string ##1\space}\immediate
      \write\@auxout{\string\bibcite{#2}{#1}}}\fi\ignorespaces}

\def\@bibitem#1{\item\hskip -3cm \hbox to 2cm
{\hfil {\footnotesize\draftcite{#1}}}\hskip 1cm
\if@filesw \immediate\write\@auxout
       {\string\bibcite{#1}{\the\value{\@listctr}}}\fi\ignorespaces}

\def\citen#1{\if@filesw \immediate\write \@auxout {\string\citation{#1}}\fi%
\@tempcntb\m@ne \let\@h@ld\relax \def\@citea{}%
\@for \@citeb:=#1\do {\@ifundefined {b@\@citeb}%
    {\@h@ld\@citea\@tempcntb\m@ne{\bf ?}%
    \@warning {Citation `\@citeb ' on page \thepage \space undefined}}%
    {\@tempcnta\@tempcntb \advance\@tempcnta\@ne
    \setbox\z@\hbox\bgroup\ifcat0\csname b@\@citeb \endcsname \relax
    \egroup \@tempcntb\number\csname b@\@citeb \endcsname \relax
    \else \egroup \@tempcntb\m@ne \fi \ifnum\@tempcnta=\@tempcntb
    \ifx\@h@ld\relax \edef \@h@ld{\@citea\csname b@\@citeb\endcsname}%
    \else \edef\@h@ld{\hbox{--}\penalty\@highpenalty
    \csname b@\@citeb\endcsname}\fi
    \else \@h@ld\@citea\csname b@\@citeb \endcsname \let\@h@ld\relax \fi}%
\def\@citea{,\penalty\@highpenalty\hskip.13em plus.13em minus.13em}}\@h@ld}
\def\@citex[#1]#2{\@cite{\citen{#2}}{#1}}%
\def\@cite#1#2{\leavevmode\unskip\ifnum\lastpenalty=\z@\penalty\@highpenalty\fi%
  \ [{\multiply\@highpenalty 3 #1%
  \if@tempswa,\penalty\@highpenalty\ #2\fi}]}   %
\makeatother 

\def\cala  {{\cal A}}
\def\calc  {{\cal C}}
\def\cald  {{\cal D}}

\def\calf  {{\cal F}}
\def\calg  {{\cal G}}
\def\calh  {{\cal H}}

\def\calm  {{\cal M}}
\def\caln  {{\cal N}}

\def\calv  {{\cal V}}

\def\dl            {\mathbb }
\def\complex       {{\dl C}}

\def\reals         {{\dl R}}
\def\zet           {{\dl Z}}
\def\Zet           {${\dl Z}$}

\def\zetpluso      {\mbox{$\zet_{\ge0}$}}

\def\alg           {algebra}
\def\Alpha         {N}
\def\auto          {automorphism}

\def\bA            {|{\cal B}^A\rangle}
\def\Bar           {\tilde}
\newcommand\baR[1] {{#1}^*}
\newcommand\Barray[2] {[\!\!{\scriptstyle\begin{array}{c}{}\\[-1.92em]
                   {\scs #1}\\[-.33em]{\scs #2}\\[-.4em] \eear}\!\!]}
\def\bc            {boundary condition}
\def\be            {\begin{equation}}
\def\bfe           {{\bf1}}
\newcommand\binom[2]{\mbox{{\large(}$\begin{array}{c}{}\\[-1.55em]\!\!\scs#1\!\!                   \!\\[-.44em]\!\!\scs#2\!\!\!\end{array}${\large)}}\,}
\def\boundk        {|{\rm B}\rangle}
\def\boundlb       {\langle{\rm B}_\Lambda|}
\def\boundlk       {|{\rm B}_\Lambda\rangle}
\def\boundlp       {|{\rm B}_{\Lambda'}\rangle}
\def\boundok       {|{\rm B}_0\rangle}
\def\boundqk       {|{\rm B}_q\rangle}
\def\branek        {|{\rm D}\rangle}
\def\brane0k       {|{\rm D}_0\rangle}
\def\braneqk       {|{\rm D}_q\rangle}

\def\CX            {M}
\def\CA            {\mbox{$\liefont C$}}
\def\cb            {chiral block}
\def\Cb            {Chiral block}
\def\cco           {{\cal O}_{\rm c}}
\def\cft           {conformal field theory}
\def\Cft           {Conformal field theory}
\def\cfts          {conformal field theories}
\def\chii          {{\raisebox{.15em}{$\chi$}}}
\def\chir          {\mbox{$\liefont W$}}
\newcommand\coi[2] {\lfloor #1\rfloor^{}_{#2}}

\def\Conetofour    {C_{\Lambda_1\Bar\Lambda_1,\Lambda_2\Bar\Lambda_2,\Lambda_3
                   \Bar\Lambda_3,\Lambda_4\Bar\Lambda_4}^{(\mu,\Bar\mu)}}
\def\corfu         {correlation function}
\def\cp            {Chan\hy Paton }
\def\crossk        {|{\rm C}\rangle}
\def\crosskd       {|{\rm C}_q\rangle_{\rm D}^{}}
\def\crosslk       {|{\rm C}_\Lambda\rangle}
\def\crossqk       {|{\rm C}_q\rangle}

\def\CTX           {M^{\rm T}}
\def\Ctilde        {\hat C}
\def\ctype         {Chan\hy Paton type}
\def\cvo           {chiral vertex operator}
\newcommand\CVO[3] {\mbox{{\Large(}$\!\!\!\begin{array}{c} {}\\[-1.49em]\scs#3\;
                   #2\\[-.30em] #1  \end{array}\!\!\!${\Large )}}}
\newcommand\CVOB[3]{\mbox{{\Large(}$\!\!\!\begin{array}{c} {}\\[-1.49em]\scs\Bar
                   #3\;\Bar #2\\[-.26em] \Bar #1 \end{array}\!\!\!${\Large )}}}
\def\dalpha        {|{\cal D}^a\rangle}
\def\dsum          {\displaystyle\sum}
\def\ee            {\end{equation}}
\def\eE            {{\rm e}}
\def\eear          {\end{array}}
\newcommand\erf[1] {(\ref{#1})}
\def\eq            {\,{=}\,}
\def\findim        {finite-dimensional}
\newcommand\Fmat[6]{{F}^{}_{\!\sss#1#2}\Barray{#4#5}{#3#6}}
\newcommand\Fmatm[6]{{F}^{-1}_{\!\sss#1#2}\Barray{#4#5}{#3#6}}
\newcommand\Frac[2]{\mbox{\large$\frac{#1}{#2}$}}
\newcommand\fraC[2]{{#1}/{#2}}
\def\frc           {fusion rule coefficient}
\def\futnot#1      {}
\def\futnote#1     {\footnote{~#1}\ }
\def\g             {{\liefont g}}
\def\gb            {{\bar\g}}
\def\gh            {{\liefont h}}

\def\GLd           {\mbox{GL$(d{,}\reals)$}}
\def\GLe           {\mbox{GL$(1{,}\reals)$}}
\def\hil           {{\cal H}}
\def\hill          {{\cal H}_\Lambda}
\def\hilo          {{\cal H}^{\sss(\om)}}
\def\hilq          {\mbox{${\cal H}_q$}}
\newcommand\hsp[1] {\mbox{\hspace{#1 em}}}
\def\hwv           {highest weight vector}
\def\hy            {$\mbox{-\hspace{-.66 mm}-}$}
\def\I             {I}
\newcommand\IA[3]  {\mbox{{\Large(}$\!\!\!\begin{array}{c} {}\\[-1.55em]\scs#3\;
                   #2\\[-.30em] #1 \end{array}\!\!\!${\Large )}}}

\def\Ic            {\I_{\rm c}}

\def\Id            {\I_{\rm d}}
\def\ii            {{\rm i}}
\def\Illl          {\CVO\Lambda\mu\nu}
\def\iN            {\,{\in}\,}
\def\infdim        {infinite-dimensional}
\def\Io            {{\cal I}_\om}

\def\irrep         {irreducible representation}
\def\Iset          {\Xi}
\def\kma           {Kac\hy Moody algebra}
\def\KZ            {Knizh\-nik\hy Za\-mo\-lod\-chi\-kov}
\def\kze           {Knizh\-nik\hy Za\-mo\-lod\-chi\-kov equation}
\long\def\labl#1   {\label{#1}\ee \ifnum\draftcontrol=1
                   \mbox{ }\\[-12 mm]\query{#1}\\[5 mm] \fi}
\def\Langle        {\langle}
\def\LAngle        {\langle}
\def\Ldots         {,...\,,}
\def\lie           {Lie algebra}
\def\liechir       {\mbox{$\liefont L$}}
\def\liefont       {\mathfrak }
\def\llb           {\mbox{\large(}}
\def\Llb           {\mbox{\Large(}}
\def\lrb           {\mbox{\large)}}
\def\Lrb           {\mbox{\Large)}}
\def\Mapsto        {\;\mapsto\;}
\def\mi            {\,{-}\,}

\def\nc            {n_{\rm c}}
\def\nE            {\,{\not=}\,}
\def\no            {n_{\rm o}}
\newcommand\nxt[1] {\\[.1em]\raisebox{.12em}{\rule{.35em}{.35em}}\hsp{.6}#1}
\def\om            {\omega}
\def\oma           {A}
\def\Oma           {(\om{,}a)}
\def\omb           {B}
\def\Om            {\Theta}
\def\one           {\mbox{\small $1\!\!$}1}
\def\onedim        {one-dimen\-sional}
\newcommand\N[3]   {\mbox{$\caln_{\!#1#2}^{\;\ #3}$}}
\newcommand\Nt[3]  {\mbox{${}_{}^{{\sss(\om)\!\!}}\caln_{\!#1#2}^{\;\ #3}$}}
\def\nontriv       {non-trivial}
\newcommand\normord[1] {\,\raisebox{.033em}{\large\bf:}#1
                   \raisebox{.033em}{\large\bf:}\,}

\def\Od            {\mbox{O$(d)$}}
\def\Oe            {\mbox{O$(1)$}}
\def\ot            {\raisebox{.07em}{$\scriptstyle\otimes$}}
\def\oT            {\,\ot\,}
\def\otimeS        {\,{\otimes}\,}
\newcommand\phb[1] {\Psi_{#1}}
\newcommand\phB[2] {\Psi^{#1}_{#2}}
\newcommand\pHB[3] {\Psi^{#1#2}_{#3}}
\newcommand\phd[1] {\varphi_{#1}}
\def\phI           {\varphi}
\newcommand\pho[1] {\phi_{#1,\Bar{#1}}}
\newcommand\phoq[1]{\phi_{#1_q,\Bar{#1}_q}}
\def\pl            {\,{+}\,}
\def\Pn            {{{\dl P}^1_{\!\!\sss(n)}}}
\def\pslz          {\mbox{PSL$(2{,}\zet)$}}
\def\qft           {quantum field theory}
\long\def\query#1{\hskip 0pt{\vadjust{\everypar={}\small\vtop to 0pt{\hbox{}%
     \vskip -13pt\rlap{\hbox to 49.0pc{\hfil{\vtop{\hsize=8pc\tolerance=6000%
     \hfuzz=.5pc\rightskip=0pt plus 3em\noindent#1}}}}\vss}}}}%
\def\Rangle        {\rangle}
\def\RAngle        {\rangle}
\newcommand\rb[3]  {R^{#1}_{#2 \Bar{#2};#3}} 
\newcommand\rB[3]  {R^{#1}_{#2 #2;#3}} 
\newcommand\RB[4]  {R^{#1}_{#2 #3;#4}}
\newcommand\rc[3]  {R^{#1}_{#2 \Bar{#2};#3}}
\def\rep           {representation}
\def\resp          {respectively}
\def\rmd           {{\rm d}}
\def\rp            {\pi}
\def\scs           {\scriptstyle}
\newcommand\sect[1]{\section{#1}\setcounter{equation}{0}}

\def\sss           {\scriptscriptstyle}
\def\sutwo         {\mbox{$\liefont{su}(2)$}}
\def\sigmac        {\sigma_{\rm c}}
\def\sigmad        {\sigma_{\rm d}}
\def\tausigma      {\theta_\om}
\def\threedim      {three-dimensional}
\def\tildeC        {{\hat C}}
\def\To            {\mapsto}
\def\tom           {t^{\sss(\om)}}
\def\ttype         {automorphism type} 
\def\Ttype         {Automorphism type} 
\def\twodim        {two-dimen\-si\-o\-nal}
\def\tya           {{\om}}
\def\tye           {{\rm id}}
\def\U             {{\sf U}}
\def\vac           {\Omega}
\def\vechil        {\hil_{\vec\Lambda}}
\def\vechilt       {\hil_{\vec{\tilde\Lambda}}}
\def\vectau        {\theta_\om^{(\vec\Lambda)}}
\newcommand\version[1] {\ifnum\draftcontrol=1 \typeout{}\typeout{#1}\typeout{}
                   \vskip3mm \centerline{\fbox{{\tt DRAFT -- #1 -- }
                   {\small\draftdate}}}
                   \vskip3mm \fi}
\def\wrt           {with respect to }
\def\wrtt          {with respect to the }
\def\wzwm          {WZW model}
\def\wzwt          {WZW theory}
\def\wzwts         {WZW theories}

\def\draftdate{\number\month/\number\day/\number\year\ \ \ \hourmin }

\global\def\draftcontrol{0}
\catcode`\@=12

\documentclass[12pt]{article} \usepackage{amssymb,amsfonts}

\begin{document}



\begin{flushright}  {~} \\[-15 mm]  {\sf hep-th/9712257} \\[1mm]
{\sf CERN-TH/97-369} \\[1 mm]
{\sf December 1997} \end{flushright}
 
\begin{center} \vskip 15mm
{\Large\bf BRANES: FROM FREE FIELDS TO}\\[4mm]
{\Large\bf GENERAL BACKGROUNDS}\\[16mm]
{\large J\"urgen Fuchs} \\[3mm]
Max-Planck-Institut f\"ur Mathematik \\[.6mm] 
Gottfried-Claren-Str.\ 26, \  D -- 53225~~Bonn \\[11mm]
{\large Christoph Schweigert} \\[3mm] CERN \\[.6mm] CH -- 1211~~Gen\`eve 23
\end{center}
\vskip 20mm
 
\begin{quote}{\bf Abstract}\\[1mm]
Motivated by recent developments in string theory, we study the structure of 
boundary conditions in arbitrary conformal field theories. A boundary
condition is specified by two types of data: first, a consistent collection of
reflection coefficients for bulk fields on the disk; and second, a choice of 
an automorphism $\omega$ of the fusion rules that preserves conformal weights. 
Non-trivial automorphisms $\omega$ correspond to D-brane configurations for 
arbitrary conformal field theories. The choice of the fusion rule automorphism 
$\omega$ amounts to fixing the dimension and certain global topological 
features of the D-brane world volume and the background gauge field on it.\\
We present evidence that for fixed choice of $\omega$ the boundary conditions 
are classified as the irreducible representations of some commutative 
associative algebra, a generalization of the fusion rule algebra. Each of 
these irreducible representations corresponds to a choice of the moduli for 
the world volume of the D-brane and the moduli of the flat connection on it.
\end{quote}
\vfill {}\begin{flushleft}  {~} \\[-3 mm]{\sf CERN-TH/97-369}
\\[1 mm]{\sf December 1997} \end{flushleft}
 
\sect{Introduction}

In this paper we investigate the structure of conformal field theories on 
(real \twodim) surfaces which have boundaries and\,/\,or are unorientable. 
These results are of relevance to various applications of such \cfts.
For example, boundary effects are of interest in the description of \twodim\ 
critical systems in statistical mechanics, the quantum Hall effect,
various impurity problems, or the Ising model with a defect line
(see e.g.\ \cite{affl8,osaf2} and the literature cited there).

Our main motivation comes, however, from string theory. Our results 
pave the way to the study of open strings and D-brane configurations not only
for backgrounds based on free field theories, but for arbitrary \cfts.
Recall \cite{GRsw} that in string theory it has been known for a long time 
that there are theories of open strings, in which the ends of the open 
strings give rise to boundaries of the string world sheet. It is also known 
that consistency of open strings requires the consideration of unoriented 
world sheets as well. In the orbifold-inspired language that has been 
proposed in \cite{sagn,hora}, the unorientable surfaces implement just
the projection on symmetric respectively anti-symmetric states.

It has been known for quite some time, too, that the low energy effective
actions of various string theories possess solitonic solutions. What has
become apparent more recently is that there is also a string perturbation 
theory for the corresponding sectors
and that this perturbation can be described using 
world sheets with boundaries ($\!\!$\cite{polc3}, see also \cite{dfpslr}). 
In the case of free strings, a Lagrangian description in terms of the string 
coordinates (Fubi\-ni\hy Ve\-ne\-zi\-ano fields) $X^i$ is available, in which 
the presence of these sectors corresponds to imposing boundary conditions on 
the $X^i$ that are more general than the usual Neumann boundary
conditions; see e.g.\ \cite{schw9} and the literature cited there.
While these considerations have 
provided non-trivial insight into the structure of string theory, they
are also subject to certain limitations. In particular, most of them 
have been restricted to BPS sectors, or can so far be formulated at a fully 
nonperturbative level (on the world sheet) only for {\em free\/} \cfts. 
(Of course, for non-free theories that possess a geometric interpretation,
one can still employ the sigma model approach, but then for many purposes
one must resort to sigma model perturbation theory.)

In this paper we present a few steps towards a deeper understanding of
interacting \cfts\ on surfaces that are relevant to open string theories. 
For brevity we will refer to \cft\ on
such world sheets which arise only in open string theories as
{\em open \cft\/}, while for \cft\ on the world sheets that already appear
for closed strings we use the term {\em closed \cft\/}. We will show in this 
paper that theories on both types of world sheets can be described in the 
same formalism. The main objective of our paper is to outline a
conceptual framework for open \cfts. Among other things, we will point out
various detailed problems that deserve further study. 
Our results put these problems in the appropriate conceptual framework and
can therefore help to initiate a programme for further research. In particular, 
we develop concepts and techniques that allow (at least in principle) to
make exact, i.e.\ in particular non-perturbative, statements beyond the BPS
sectors of string theories based on interacting \cfts.

A more specific goal of our work is to gain insight into the description and
the structure of boundary conditions. 
In the geometric description of free strings via the 
Fubi\-ni\hy Ve\-ne\-zi\-ano fields $X^i$ (which are not proper conformal 
fields), D-branes are submanifolds of the target space, along with
some additional structure. More precisely, such a D-brane is characterized
by its dimension, its world volume and a vector bundle on it. We will give
analogues of all these data for arbitrary \cft\ backgrounds.
Briefly, our basic observation in this direction is that for an arbitrary
\cft\ there exists a structure that, when specialized to
the \cft\ of free bosons, reproduces the various possible topologies of 
the D-brane. This structure is provided by certain
automorphisms $\om$ of the fusion rules, in much the
same way as fusion rule automorphisms enter the classification of modular
invariant torus partition functions in the case of closed \cft.

Our next result concerns the finer structure of these boundary
conditions. We exhibit certain algebras which provide
a tool to classify the various possible boundary conditions. We present evidence
that such a `classifying algebra' is a commutative associative
\alg\ that generalizes the fusion rule \alg, 
that for fixed choice of $\om$ the boundary conditions are in
one-to-one correspondence to its irreducible representations.
Each of these \irrep s corresponds to a choice of the moduli for the world 
volume of the D-brane and the moduli of the flat connection on it.

The plan of this article is as follows. Our basic aim is to arrive at a better
understanding of open \cft. 
Open \cft\ has been studied for quite some time both in its own right 
\cite{card9,card10,ishi,cale,lewe3,prss3,sasT2} and in relation to strings
\cite{ales,alam,clny3,kols,dpfhls,bisa2,bips,bips2,fips,grwa,prss,prss2,abpss,%
fpslr,bidc,reSC,dfpslr}.
To some extent we can build on these existing discussions, but 
for a more comprehensive picture also some further ingredients are needed.  
In fact, it turns out that our first task is to
generalize the very concept of a \cft\ to the case of
real \twodim\ surfaces which are allowed to possess boundaries and\,/\,or to be
unorientable. In particular, a better understanding of open \cft\ 
in part requires a rather detailed discussion of several features of closed
\cft. Accordingly, in section 2 we first review the structure of \cft\ on a 
closed and orientable surface from a somewhat unorthodox point of view that 
concentrates on the relationship between `chiral' and `full' \cft\ and as a
consequence has the advantage 
that it generalizes in a straightforward manner to the open case. 
Those readers who expect that they can dispense of the information given
there altogether may proceed directly to section 3,
and later consult section 2 when necessary.
  
Afterwards, in section 3, we turn to the discussion of open \cft.
In subsection \ref{soco} we introduce the oriented cover of the world sheet
of an open \cft. Bulk fields then correspond to a suitable product of two
\cvo s on the oriented cover, as described in subsection \ref{sbul}.
In subsection \ref{scros} we study one-point chiral blocks for the situation
where the world sheet is the crosscap. Subsection \ref{sttyp} deals with
the relation between the two chiral labels $\Lambda$ and $\Bar\Lambda$
of a bulk field $\pho\Lambda$; we argue that $\Bar\Lambda\eq\om(\Lambda)$
with $\om$ an \auto\ of the fusion rules that preserves conformal weights,
and discuss the implementation of this \auto\ on \cb s and at the level of 
the operator formalism. This allows in particular to establish a formalism
for branes in arbitrary \cfts; the specific case of free bosons, for which
one recovers the known ordinary D-branes, is treated in subsubsections 
\ref{sfrb} and \ref{scfb}. The concepts of boundary conditions and boundary 
fields are introduced in subsection \ref{sbound}, which allows us in 
particular to study, in subsection \ref{sdisk}, \cb s on the disk. We are then 
in a position to proceed to the stage of full \cft, which is done in
subsection \ref{sful}; in particular the concept of reflection coefficients
is introduced there. In subsection \ref{scla} we address the issue of
a classification of all possible boundary conditions (for fixed choice of the
\auto\ $\om$); this leads us to the concept of a `classifying \alg', which
we illustrate by various examples.

In section 4 we briefly add some remarks concerning the possible
application of our results to string theory. We end in section 5 by
outlining further lines of research, both for \cft\ and for string theory.

Finally we mention that we have written this article with an eye towards 
applications in string theory. Accordingly, at several instances we
streamline the arguments by leaving aside certain aspects that are presumably
irrelevant in string theory. For example, we do not display
explicitly the Weyl anomaly because it cancels out
in critical string theory. We also suppress most mathematical issues
that concern the topology on the space of physical states and
domain questions for unbounded operators which act on that space.
Since we are able to analyze open \cft\ in a manner completely analogous
to closed \cft, these matters can be addressed by the same methods
that were developed in the latter context, see for example \cite{fefk,fefk3},
and there is no reason to expect any further complications in the open case.

\sect{Closed \cft\ revisited}

\subsection{Chiral versus full \cft}

\subsubsection{Orientability versus orientedness}

In quantum physics, one faces quite often the following situation.
One is given a collection of data of which one expects that they should be 
sufficient to characterize some quantum theory completely (this expectation
can e.g.\ be based on the observation that these data would
already suffice to specify a corresponding classical theory), but closer
inspection reveals that in order to be able to give a complete definition of
the quantum theory in fact some additional auxiliary structure is needed.
In such a situation, one has to make sure that in the end this auxiliary 
structure can again
be eliminated without affecting the observable predictions of the theory.\,%
\futnote{When this is not possible, then one speaks of an anomaly.}
A nice example of this phenomenon is provided by `topological' field
theories. One would like to define such a theory on any differentiable manifold
$M$ by only using the differentiable structure of $M$. But in order to 
formulate the theory as a quantum field theory (e.g.\ via a path integral), 
one actually needs to endow $M$ with the auxiliary structure of a metric.

As it turns out, this pattern is also realized in the situation of our interest.
Namely, our goal
is to set up a \twodim\ \cft, or more precisely, a \cft\ on some (real) \twodim\
differentiable manifold $C$. But in order to achieve this goal, it is in fact
necessary to consider manifolds that possess the structure of a complex curve,
i.e.\ a {\em complex\/} manifold of dimension one. 
The origin of this requirement is that we have to impose invariance under local
conformal transformations. Technically, this means that one must endow the
(real) \twodim\ manifold with a {\em conformal structure\/} -- that is, with
an equivalence class of metrics that are related by local rescalings.
In two dimensions, a conformal structure is, in turn, 
equivalent to a {\em complex structure\/}.
It is important to realize that the choice of a complex 
structure requires in particular the choice of an orientation.
Now even when the surface $C$ is orient{\em able\/}, it does {\em not\/}
come as an orient{\em ed\/} surface; in other words, none of the two possible 
orientations is preferred over the other. It follows that once we have achieved
the construction of a \cft\ on an oriented surface, we finally have to 
eliminate any dependence on the chosen orientation.

\subsubsection{The two stages of \cft}

{}From these observations
we conclude in particular that in the study of \cft\ there are two distinct
conceptual levels that should better be carefully distinguished: 
\nxt \Cft\ on a complex curve $\Ctilde$. We will refer to this stage as the 
     {\em chiral \cft\/}.\,%
\futnote{It is worth mentioning that it is usually this kind of structure
that mathematicians refer to when they talk about `\cft', see e.g.\ 
\cite{tska3,tsuy,falt,fesv2,beLa}.}
\nxt \Cft\ on an unoriented real \twodim\ manifold $C$.
     We call this stage the {\em full \cft\/}.\\[.1em]
Furthermore, we proceed from the former to the latter by eliminating the choice
of an orientation.  Let us stress that our approach to this issue differs from 
the usual description that one finds in the literature. Conventionally, one 
imagines that one could study the theory directly on a real \twodim\ manifold. 
Now in many cases, in which one employs a path integral formulation of the 
theory, one observes that the space of solutions to the corresponding classical
field equations factorizes. This is then taken as a motivation to `split' the 
theory into two `chiral halves', each of which is afterwards studied as 
an independent \cft\
on a complex curve. This recipe is sometimes expressed by saying that one can 
`treat $z$ and $\bar z$ as independent complex variables'. In contrast, for us 
the starting point is the chiral theory on the curve $\Ctilde$, which can and 
should be studied in its own right, and the \cft\ on the real \twodim\ 
manifold is obtained only at a later stage
by eliminating the orientation dependence. In our
opinion this provides in particular a more natural rationale for the
emergence of `left-' and `right-movers'. Moreover, as we will see
it has the additional benefit that it works for the open case as well.
(Also note in this context that all our considerations will be at a fully 
non-perturbative level for the world sheet theory; in particular
we do not have to assume that the \cfts\ under consideration possess any
Lagrangian formulation 
so that there need not be any analogue of field equations.)

To proceed, we will first have to exhibit the central pieces of structure 
that are present for chiral \resp\ full \cft, and also those that arise 
in string theory. Somewhat unexpectedly, a careful discussion of these topics 
provides us with new insight already for the case of closed \cft, 
namely concerning the relation between the chiral and the full level.
For instance, the role of fusion rule automorphisms for the classification of 
modular invariants can be easily understood via this relationship. Moreover, 
it will turn out that once we have established a suitable -- not entirely
conventional -- framework for closed \cft, the extension to the
open case, both for \cft\ proper and for string theory, is much more
straightforward than in more conventional approaches.
In this framework the basic ingredient is the system of \cb s. We
analyze these quantities from a point of view that does not presuppose the
existence of an operator formalism, neither for the full nor just for the
chiral \cft. For technical reasons, some aspects of this approach
(in particular the description of \cb s as so-called co-invariants) can so far 
be made fully explicit only for \wzwts. But it is generally expected (see e.g.\ 
\cite{mose4}) that the relevant
structures that are available in the WZW case are merely specific realizations 
of general structures that are indeed present in any arbitrary \cft.

\subsubsection{Perturbative string theory}

On top of these two conceptual stages of \cft, for the applications we have in 
mind there is a third, the one of (perturbative) string theory. At this 
additional stage the 
guiding principle is to get rid of {\em all\/} properties of the world sheet $C$
while still keeping information about the \qft\ that was defined on
the world sheet. To implement this principle one first eliminates 
the Virasoro algebra (or one of its (super-)extensions) by taking the relevant 
(semi-infinite) cohomology. Next the choice of a conformal structure is
eliminated by performing an integral over the moduli space of complex 
structures. And finally one eliminates the choice of topology of the world 
sheet by summing over all possible topologies, where the sum is weighted by a 
factor of $(\gamma_S)^{-\chi}_{}$, with $\gamma_S$ the string coupling 
constant and $\chi$ the Euler number of the world sheet.
(Thus in particular, when one uses the \corfu s of the \cft\ for the computation
of string scattering amplitudes, the latter will always come
combined with the relevant moduli integrals.)

\subsection{Chiral \cft}

We start by discussing {\em chiral\/} \cft. That is, we work on some
manifold $\tildeC$ that has the structure of a complex curve.
Technically, $\tildeC$ is an algebraic curve over the complex numbers $\complex$
that is complete and reduced and whose singularities are at worst ordinary 
double points. (The inclusion of the singular curves serves to compactify 
all moduli spaces $\calm_{g,n}$ that will enter our investigations.)

\subsubsection{Chiral sectors}

The main structural data of a chiral \cft\ are provided by the {\em system of 
\cb s\/}. It is worth stressing that a priori there may well exist 
such systems which cannot be constructed in an operator formalism, 
i.e.\ systems for which the \cb s cannot be interpreted directly as
matrix elements of products of (chiral) vertex operators. 
But since the connection to more heuristic considerations is usually made
via an operator formalism, in the present and next subsubsection we
first list those ingredients which occur in an operator interpretation of 
\cb s (compare e.g.\ \cite{fefk3,mose3,Algs,Mose,Sche2',jf27}).
\nxt 
The observables form some infinite-dimensional associative \alg\ \chir, 
called the {\em chiral symmetry algebra\/}.  \chir\ is \Zet-graded, it is 
a *-\alg\ (i.e.\ is endowed with an involutive anti-\auto), and it
contains (the enveloping \alg\ of) the Virasoro \alg\ as a sub\alg.
The grading is provided by the zero mode $L_0$ of the Virasoro \alg.
\\
Often instead of this associative \alg\ one can equivalently consider
the Lie \alg\ \liechir\ whose Lie bracket is given by the commutator \wrt 
the associative product. But in general there is no guarantee that this
already captures all features of the associative \alg\ \chir.
\nxt
The space of states of the chiral theory is the direct sum 
  \be  \hil = \bigoplus_{\Lambda\in\Iset}\hil_\Lambda\,,  \labl{hil}
where each of the (chiral) {\em sectors\/} $\hil_\Lambda$ is some \infdim\ 
vector space, whose precise structure depends on the framework one chooses. In 
a representation theoretic approach, it is natural to assume that the space 
$\hil_\Lambda$ has a 
gradation over the integers that is compatible with the $\zet$-grading of the 
chiral algebra. Moreover, all subspaces of definite grade should be \findim,
and their dimensions should grow less than exponentially.
The latter condition ensures that the Virasoro-specialized character 
  \be \chii_\Lambda (\tau) := {\rm tr}_{\hil_\Lambda}^{} \eE^{2\pi\ii
  \tau(L_0-c/24)}_{}  \labl{cha}
converges for any $\tau$ in the complex upper half-plane. 
\nxt
We will further 
assume that the sectors carry a scalar product such that the action of the 
chiral algebra \chir\ is unitary, i.e.\ we restrict ourselves to unitary \cfts.
In a more field theoretical spirit, one would further require that the sectors 
are endowed with the additional structure of a Hilbert space. Also,
when the Lie \alg\ \liechir\ associated to the chiral \alg\ possesses 
a triangular  decomposition, the sectors $\hil_\Lambda$ will usually be
irreducible highest weight modules over \liechir.
\nxt
We will refer to the elements $\Lambda$ of the index set $\Iset$ that
appears in \erf{hil} as `weights' of the sectors, or also as `sector labels'.
When the set $\Iset$ of sector labels
is finite, then the \cft\ is called {\em rational\/}.
\nxt
There is a notion of {\em fusion product\/} \cite{fefk3,gabe3} 
which associates to any pair $\hil_\lambda,\,\hil_\mu$ of sectors a 
direct sum $\bigoplus_{\nu\in\Iset} \N\lambda\mu\nu \hil_\nu$ of sectors. The 
fusion product preserves in particular the eigenvalues of central charges, and 
hence does not coincide with the ordinary tensor product of \liechir-modules.
\nxt
The non-negative integers $\N\lambda\mu\nu$ are known as the {\em fusion rule 
coefficients\/} of the theory. Furthermore, there is a distinguished sector 
$\hil_\vac$, the {\em vacuum sector\/}, which concerning fusion plays the role 
of a unit element, i.e.\ $\N\vac\lambda\mu\eq\delta_\lambda^{\;\mu}$.
Moreover, $\N\lambda\mu\vac\eq\delta_{\lambda,\mu^+}$, where 
  \be  \mu\mapsto\mu^+  \labl{mm}
is a permutation that preserves the fusion rules as well as the conformal 
weights; the sector $\hil_{\mu^+}$ is called {\em conjugate\/} (or also charge
conjugate) to $\hil_\mu$.

\subsubsection{Chiral vertex operators}

In the usual language, which is borrowed from ordinary quantum field theory, 
one associates to each sector $\hil_\Lambda$
a primary `field' and its descendants. However, such an `operator formalism'
must be introduced with great care. For chiral \cfts, the
relevant concept is the one of {\em\cvo s\/} \cite{tska3,resC3,mose3}, which 
constitute intertwiners for the fusion product \cite{fefk3,zhu2}.
Technically, a \cvo\ for the sector $\hil_\Lambda$ is a linear map\,%
\futnote{Incidentally, this is one instance where we should worry about 
the completion of modules. For dealing with operators that make sense for a
sufficiently large set of $z$-values, in this formula at least the range 
sector $\hil_\nu$ must be understood as being suitably completed.}
  \be \begin{array}{ll} {\cal V}_\Lambda:\, 
  &\ \hil_\Lambda \;\to\; 
     z_{}^{\Delta_\nu-\Delta_\Lambda-\Delta_\mu}\,
  {\rm Hom}(\hil_\mu{,}\hil_\nu)[[z,z^{-1}]]  \\{}\\[-.8em]
  & |\psi_\lambda\rangle \;\mapsto\; {\cal V}_\Lambda(|\psi_\lambda\rangle)
  =: \CVO\Lambda\mu\nu(\psi_\lambda;z)  \end{array} \labl{cvo}
for a fixed choice of sectors $\hil_\mu$ (the `source' sector) and $\hil_\nu$
(the `range' sector) that possesses certain intertwining properties
for the chiral \alg\ \chir, or more precisely, for the action of \chir\ on
the fusion product of $\hil_\lambda$ and $\hil_\mu$ and on $\hil_\nu$, \resp.
For a given triple $\Lambda,\mu,\nu$ there can in general exist several
independent maps of this type; the dimension of the space of such maps is just 
given by the fusion rule $\N\lambda\mu\nu$. Throughout
the paper we suppress the corresponding multiplicity labels.
\nxt
Usually we will have to deal only with {\em primary\/}
\cvo s, by which one means the image of the \hwv\
$|\psi_\Lambda\rangle$ of $\hil_\Lambda$; we denote this linear map by 
  \be  \CVO\Lambda\mu\nu(z) \equiv \CVO\Lambda\mu\nu(\psi_\Lambda;z) :=
  {\cal V}_\Lambda(|\psi_\Lambda\rangle) \,.  \labl{pc}
Moreover, we will follow the common practice to abbreviate this quantity by 
the symbol 
  \be  \phI_\Lambda^{}(z) \equiv \CVO\Lambda**(z) \,,  \labl{phi}
i.e.\ suppress the source and range labels, whenever 
this makes the formulas more intelligible.
\\
In the case of free bosons $X^i$, the fusion rules are abelian, i.e.\
$\N\lambda\mu\nu\eq\delta_{\nu,\lambda+\mu}$, and the primary
\cvo s are nothing but the usual abelian vertex operators \cite{GRsw}, 
which can be written as normal ordered exponentials 
 \be  \phI_\lambda^{}(z) = \normord{\eE^{\ii\lambda\cdot X(z)}} \labl{ver}
of the free (chiral) boson fields.
\nxt
At genus zero, products of $m$ \cvo s $\phi_{\Lambda_i}(z_i)$ can be defined 
when the variables $z_i$ lie in the subset $\{ (z_i) \,{\mid}\, |z_m|\,{>}\, 
\cdots \,{>}\,|z_2|\,{>}\,|z_1|,\; {\rm arg}(z_i)\,{>}\,0 \}$ of $\complex^m_{}
$, and via the intertwining properties the domain of definition can be extended 
to the image of this set under projective transformations, which is the 
universal covering of $\complex^m_{}{\setminus}\{ z_i\eq z_j\;{\rm for\; 
some}\;$\linebreak[0]$i\nE j\}$ \cite{fefk3}. In rational theories, one 
is actually dealing with a finite covering. This shows in particular that 
in a rational theory the variable $z$ in \erf{cvo} must be interpreted as  
$z\eq w^\ell$ with some $\ell\iN\zet$ and $w$ a (quasi-)global coordinate on 
this finite covering. Very 
roughly, one can think of $z_i$ as the value that the quasi-global coordinate
function $z$ on ${\dl P}^1$ takes at the insertion point $p_i$ of the \cvo.\,%
\futnote{Note that we must distinguish carefully between the point $p_i$
and the coordinate value $z_i\eq z(p_i)$. This is nicely illustrated by
the appearance of powers $z_i^n$ below, which make sense for coordinates, while 
of course there does not exist any concept of multiplying the points $p_i$.}
An analogous interpretation is definitely no longer possible at higher genus, 
where there is no analogue of the quasi-global coordinate $z$.
\nxt
It is commonly expected that the fusion product and the \cvo s
can be constructed in a mathematically rigorous manner via the representation 
theory of vertex operator algebras \cite{FRlm,KAc4,hule3,hule5}, where the 
indeterminate $z$ in \erf{cvo} is regarded as a formal variable. It is, 
however, not clear to us whether that settles the problems that potentially 
arise from the fact that the sectors $\hil_\lambda$ are not complete (in the 
norm topology) and that the \cvo s are unbounded operators. This remark applies
likewise to the supposed operator formalism for open \cfts, where the situation
is complicated by the fact that one also has to take care of the various 
possible boundary conditions.

\subsubsection{\Cb s}\label{scb}

Even though the \cvo s are quite directly accessible to field theoretic 
intuition, the more important structure in chiral \cft\ is actually provided 
by the {\em\cb s\/}. (This can e.g.\ be inferred from the fact that even in the 
operator formalism the \cb s are the prime quantities of interest as soon as 
it comes to concrete calculations.) In the operator formalism, \cb s are easily 
constructed for arbitrary \cfts\ for curves $C$ of genus 0 or 1.
At genus $0$, where the concept of radial ordering makes sense, 
one can in particular multiply \cvo s in a well defined manner;
the \cb s are then given by the expectation values 
  \be V_0(\vec\Lambda;\vec\mu) \equiv \langle
  \phI_{\Lambda_1}(z_1)\, \phI_{\Lambda_2}(z_2)\cdots \phI_{\Lambda_{n-1}}
  (z_{n-1})\, \phI_{\Lambda_n}(z_n)\rangle_{}^{(\vec\mu)} \,, \labl{cb0}
of such products of primary \cvo s; here 
$\langle\cdots\rangle\equiv\langle\vac|\cdots|\vac\rangle$ denotes the
vacuum expectation value, while the additional label $\vec\mu$ indicates the
chosen collection of `intermediate states' including possible multiplicities.\,%
\futnote{In the special case that the \frc s satisfy $\sum_{\nu\in\Iset}
\N\lambda\mu\nu\eq1$ for all $\lambda,\mu\iN\Iset$ (e.g.\ for free bosons,
where one has $\N\lambda\mu\nu\eq\delta^{}_{\nu,\lambda+\mu}$), only a single
such choice is possible, but generically there are 
$\sum_{\mu_1,\mu_2,...\in\Iset} \N{\Lambda_1}{\Lambda_2}{\ \mu_1}
\N{\mu_1}{\Lambda_3}{\ \mu_2}\cdots \N{\mu_{n-3}}{\Lambda_{n-1} \Lambda_n}{}$
many distinct possibilities.}
(In a pictorial description of the blocks by graphs with trivalent vertices, 
the intermediate states correspond to the labels of the internal lines,
while possible multiplicities correspond to labels of the vertices; for
a review of this pictorial \rep\ see e.g.\ section 1.2 of \cite{sorg}).
Similarly, for a curve of genus $1$ with modular parameter $\tau$
the \cb s can be obtained as suitably weighted traces of products of \cvo s,
  \be  V_{1;\nu}(\vec\Lambda;\vec\mu) \equiv
  {\rm tr}_{\hil_\nu}^{}\llb \eE^{2\pi\ii \tau(L_0-c/24)}_{} \,
  \phI_{\Lambda_1}(z_1)\cdots \phI_{\Lambda_n}(z_n)\lrb_{}^{(\vec\mu)} \,, 
  \labl{cb1}
where $\vec\mu$ has an analogous meaning as at genus zero and $\nu$ corresponds
to yet another internal line that closes the corresponding graph to a `1-loop 
diagram'; a special example is provided by the characters \erf{cha}, which can 
be interpreted as zero-point blocks on the torus,
$\chii_\lambda \equiv V_{1}(\emptyset;\lambda)$.

In principle, via the implementation of suitable factorization rules 
it should be possible to establish an operator formalism that allows for
a definition of \cb s for higher genus surfaces similarly as in \erf{cb0} and
\erf{cb1}. Unfortunately, so far such an operator formalism
has not been worked out for general interacting conformal field theories.
On the other hand, such a formulation is already available to some extent 
for free field theories, or more specifically, mainly for free bosons and for
$b$-$c$-systems. These constructions are either based on a
path integral formulation \cite{vafa1,ismo,algr,abmnv} 
or employ the \rep\ theory of Krichever\hy Novikov type \alg s \cite{boto}.
Some specific non-free theories have been
analyzed as well, namely orbifolds of free boson theories \cite{divv2} and free 
bosons with a background charge \cite{dpfhls}. In fact, the results obtained 
in the latter case (which are based on the earlier work \cite{dfls,pezz})
might provide guiding principles for the extension to more complicated \cfts.
Some other results that are relevant to the case of general \cfts\ can be
found in \cite{frki}; they are based on the sewing prescriptions of 
\cite{sono,sono2}.

For our purposes, the proper way to think about the \cb s is as the space 
of solutions to certain algebraic and differential equations which are known as 
the {\em Ward identities\/} of the theory. 
The Ward identities for \cb s form a collection
of algebraic and differential equations that express the symmetries of the 
theory. In the operator formalism they arise by combining the special 
properties of the vacuum state (e.g.\ that it is annihilated by the Virasoro 
modes $L_m$ with $m\,{\ge}\,{-}1$) with the intertwining properties of 
the \cvo s. It is worth stressing that here we speak of {\em all\/} Ward 
identities, not only those particularly simple ones (such as the projective 
Ward identities, which correspond to the modes $L_0$ and $L_{\pm1}$ of the 
Virasoro \alg) that are obtained by considering those modes
of the chiral algebra which annihilate both the `in' and the `out' vacuum
state. (The \cb s also satisfy a collection of identities
that are commonly called null vector equations. In the present description
all the information contained in those null vector equations is already
implemented by the Ward identities together with
fact that the sectors are irreducible modules.)
However, except for the latter special cases, it proves to be a difficult
task to write out the Ward identities explicitly. This suggests to look
for an alternative formulation of the Ward identities that does not make use 
of the concept of \cvo s and that, ideally, allows to combine all Ward 
identities into a single prescription. As it turns out,
one can make this idea concrete in the special case of \wzwts. 
The relevant structure turns out to be given by what is known as
the {\em block algebra\/}, which provides a `global' version
of the chiral algebra that is only defined with reference
to some local coordinate. As will be explained in some detail in subsection 
\ref{swzw}, the (co-)\,invariants \wrtt block \alg\ indeed implement all 
the Ward identities of the \wzwt.

\subsubsection{Bundles of \cb s}\label{sbcb}

By now, we have succeeded in associating to each point 
$(g;p_1,p_2 \Ldots p_n) \equiv (g;\vec p)$ in the 
moduli space $\calm_{g,n}$ of complex curves of genus $g$ with $n$ distinct 
marked smooth points $p_1,p_2 \Ldots p_n$ some complex vector space 
$V \equiv V_{\Ctilde_g}(\vec p,\vec\lambda)$, the space of \cb s. 
It is expected that these vector spaces
$V_{\Ctilde_g}(\vec p,\vec\lambda)$ for all $(g;\vec p)\iN\calm_{g,n}$
fit together to the total space of a vector bundle\,%
\futnote{Actually, since the moduli spaces $\calm_{g,n}$ have singularities, 
one is dealing with locally free sheaves instead of vector bundles.}
over $\calm_{g,n}$, the bundle $\calv \equiv \calv(\vec\lambda)$ of \cb s; 
in the WZW case this can be proven rigorously ($\!\!$\cite{tsuy,Ueno}, for a 
review see \cite{sorg}). More precisely, for any finite sequence
$\vec\lambda\equiv(\lambda_1,\lambda_2 \Ldots \lambda_n)$
of $n$ sector labels $\lambda_1,\lambda_2 \Ldots \lambda_n\iN\Iset$
of a rational chiral \cft, the solutions to the Ward identities provide us 
with a vector bundle $\calv$ of finite rank
  \be  {\rm rank}_\complex^{} V_{\Ctilde_g}(\vec p,\vec\lambda)
  =: \caln_g(\vec\lambda) \labl{ngl}
over each of the moduli spaces $\calm_{g,n}$.

This system of vector bundles has a number of highly non-trivial properties.
First, we assume that each of the vector bundles $\calv(\vec\lambda)$ comes
equipped with a projectively flat unitary connection, the
{\em\KZ\ connection\/}. The existence of a projectively flat connection has 
been shown rigorously \cite{tsuy,Ueno} for any genus in the case of \wzwts. 
(A generalization for genus zero is discussed in \cite{alrs}.)
The existence of this connection
motivates the habit (that we will also follow) to use the term \cb\ not only for
an element of the fiber over a specific point of $\calm_{g,n}$, but also 
in a closely related, but still conceptually different meaning, namely as some 
definite horizontal section of the bundle $\calv$.

Moreover, we will assume that the system of blocks obeys so-called
{\em factorization rules\/} \cite{frsh,tsuy}. They relate the \cb s
on a singular curve and the blocks on its {\em normalization\/} for which the 
singularity is resolved. Suppose e.g.\ that the curve $\tildeC$ is singular 
and that the singularity is an ordinary double point at $q\iN\tildeC$. 
In the normalization $\tilde C$ of $\tildeC$ two points $q_1,q_2$ of
$\tilde C$ lie over the singular point $q$. Then the factorization rule reads
  \be  V_{\tildeC}(\vec p\,{\cup}\{q\},(\vec\lambda,\vac)) =
  \bigoplus_{\mu\in\Iset} V_{\tilde C}(\vec p\,{\cup}\{q_1,q_2\},\vec\lambda\,
  {\cup}\{\mu{,}\mu^+\}) \,. \labl{fac1}
It can happen that the normalization has two different connected components.
This is e.g.\ the case when one pinches a curve $\tildeC$ between two sets 
$\vec p$ and $\vec p\,'$ of points in such a manner that it becomes singular. 
The normalization $\tilde C$ of $\tildeC$ has in this case two connected 
components $\tilde C_1$ and $\tilde C_2$, with the lift of $\vec p$ in 
$\tilde C_1$ and the lift of $\vec p\,'$ in $\tilde C_2$. As we will see 
in subsection \ref{sbib},
the blocks for a curve which is not connected are given by the tensor
products of blocks for the different connected components. As a consequence,
in this special situation the factorization rule \erf{fac1} reads 
  \be V_C(\vec p\,{\cup}\vec p\,',\vec\lambda\,{\cup}\vec\lambda') =
  \bigoplus_{\mu\in\Iset}
  V_{\tilde C_1}(\vec p\,{\cup}\{q_1\},\vec\lambda\,{\cup}\{\mu\})  \otimes
  V_{\tilde C_2}(\vec p\,'\,{\cup}\{q_2\},\vec\lambda\,{\cup}\{\mu^+\}) \,.
  \labl{fac2}
The \KZ\ connection and factorization rules have deep consequences 
because they provide relations between chiral \cfts\ 
on {\em different\/} curves $\tildeC$. As a consequence, they allow us to
speak in a meaningful way about the `same' chiral \cft\ on different curves. 
(In contrast, to speak about the `same \qft' on two arbitrary space-times 
$M_1$ and $M_2$ typically does not make any sense at all.)
  
The factorization rules \erf{fac1} and \erf{fac2} also have 
evident consequences on the fusion rules, i.e.\ the ranks of the vector bundles
of \cb s. More precisely, the integers 
$\caln_g(\vec\lambda)$ \erf{ngl} inherit evident properties from the 
requirements \erf{fac1} and \erf{fac2}; in particular, we have
  \be  \caln_g(\vec\lambda) = \sum_{\mu\in\Iset} \caln_{g-1}(\vec\lambda\,
  {\cup}\{\mu{,}\mu^+\})\, . \ee

There exists closed formulas, the so-called Verlinde \cite{verl2} formulas,
for the dimensions $\caln_g(\vec\lambda)$, which are automatically compatible
with the factorization rules. To explain their origin, we note that
the fundamental group of the moduli space of non-singular tori is the modular
group \pslz; it can be written as the complex upper half plane,
parametrized by $\tau$, modulo the action of \pslz, acting as
  \be \tau \mapsto \frac{a\tau+b}{c\tau+d} \, . \ee
The existence of a unitary \KZ\ connection implies that the space of 
zero-point blocks on the torus, i.e.\ the characters, carries a projective 
unitary representation of the modular group (compare also \cite{zhu3}).
The unitary symmetric matrix $S$ 
that describes the transformation of the characters under the transformation
$\tau\mapsto -1/\tau$ is of particular importance. 
Another consequence of the existence of the \KZ\ connection is that 
 -- as is already implicit in the chosen notation --
the dimension of the space of \cb s does not depend on the precise
choice of the moduli, but only on the genus $g$ of the curve and the type 
$\vec\lambda$ of insertions.\,%
\futnote{Actually, this follows already from the fact that the sheaf of
\cb s is locally free.}
The Verlinde conjecture states that for every rational \cft\ the rank of the 
vector bundle of \cb s on a curve of genus $g$ with sectors $\lambda_1 \Ldots 
\lambda_n$ as insertions is given by the expression
  \be  \caln_g(\vec\lambda) = \sum_{\mu \in\Iset}
  |S_{\mu,\vac}|^{2-2g}_{}
  \prod_{l=1}^n \frac{S_{\lambda_l,\mu}}{S_{\vac,\mu}}  \,.  \labl{vefo}

\subsection{WZW blocks as co-invariants}\label{swzw}

\subsubsection{WZW theories}

We have noted above that the description of blocks in terms of solutions to 
the Ward identities is not completely trivial. However, there is one subclass 
of \cfts, namely the \wzwts, for which the \cb s can be obtained in that 
framework in very concrete terms, such that e.g.\ the Verlinde conjecture 
\erf{vefo} can be proven rigorously \cite{beLa,falt,sorg}.
(Since this class actually serves as the starting point for obtaining many 
other \cfts, such as coset models, it is to be expected that much of the
structure that we will describe for the WZW case can be generalized.)

For these models the underlying Lie \alg\ \liechir\ is an affine \kma\ $\g$,
which is generated over $\complex$ by modes $J^a_n$ with $n\iN\zet$ and by a 
central element $K$, with Lie brackets
  \be [J^a_n, J^b_m] = \sum_cf^{ab}_{\;\ c}\, J^c_{n+m} + K \kappa^{ab}\,n\,
  \delta_{n+m,0} \,. \ee
Here $f^{ab}_{\;\ c}$ and $\kappa^{ab}$ are the structure constants and
Killing form, \resp, of the horizontal subalgebra of $\g$, i.e.\ of the
\findim\ simple \lie\ $\gb$ that is spanned by the zero modes $J^a_0$. 
For our purposes, it is convenient to allow also for 
the case that all structure constants vanish, $f^{ab}_{\;\ c} \equiv 0$. 
The corresponding models describe the \cft\ of free bosons; the modes $J^a_n$ 
are then the modes of the abelian currents $J^a(z)\equiv\ii\partial X^a(z)$ 
and span the Heisenberg \alg\ $\hat{\liefont u}(1)$ in place of an affine 
\kma\ $\g$. (As a side remark, we mention
that the system of \cb s for \wzwts\ is most intimately related to the space 
of states of the \threedim\ Chern\hy Simons theory that is based on the 
simple compact connected and simply connected Lie group whose \lie\ is the
real form of $\gb$ \cite{witt27,frki}.) For a \wzwt\
the index set $\Iset$ consists of all weights $\Lambda$ of $\g$ that are
integrable at some fixed value $k$ (the level) of the canonical central element 
$K\iN\g$; to each $\Lambda$ there is associated a unitarizable irreducible 
highest weight module $\hil_\Lambda$, which constitutes the space of states
in the corresponding sector. Given the level $k$, a highest weight $\Lambda$ 
is already determined by its horizontal part, i.e.\ by the weight 
$\bar\Lambda$ \wrt the subalgebra $\gb$.
The highest weight for the vacuum sector is given 
by $k$ times the zeroth fundamental weight of $\g$, $\vac\eq k\Lambda_{(0)}$,
and the conjugate $\Lambda^+$ of $\Lambda$ is the unique weight that has the
same level $k$ as $\Lambda$ and whose horizontal part is the conjugate
(as $\gb$-weight) of $\bar\Lambda$.

\subsubsection{Global symmetries: the block algebra}

Let us describe in some detail how one can characterize the \cb s of a \wzwt.
To keep the discussion as elementary as possible, we take $C$ to be the
complex projective space ${\dl P}^1\equiv{\dl P}\complex$\,, which we
represent as the complex $z$-plane plus a point at infinity. Given $n$ 
(distinct) points $p_i$ on ${\dl P}^1$, to which for brevity we still refer as 
the insertion points even in the absence of an operator formalism,\,%
\futnote{In mathematics, also the term `parabolic points' is common.}
and the corresponding $n$ sectors labelled by
integrable weights $\Lambda_i$ of $\g$, we consider the tensor product
  \be  \vechil:=\hil_{\Lambda_1}\otimeS \hil_{\Lambda_2}\otimeS\cdots\otimeS
  \hil_{\Lambda_n}  \ee
of the $\g$-modules $\hil_{\Lambda_i}$. {}From this big vector space we 
obtain the blocks by imposing the Ward identities. These identities should 
constitute the global realization of the symmetries of the theory and
involve states `inserted' at different points. As a consequence, we need
a new algebra that encodes these symmetries. In the case of \wzwts\ such
an algebra is readily available: one takes the algebra of all $\gb$-valued 
holomorphic functions on $\tildeC$, with singularities not worse 
than poles of finite order at the insertion points $p_i$. Note that here we 
use the concept of holomorphic function, which is well-defined only once we 
have chosen a complex structure on the manifold $\tildeC$.

The space of $\gb$-valued functions actually forms a Lie algebra; the Lie 
bracket is just the 
commutator \wrt the associative product that is given by pointwise 
multiplication. We call this \alg\ the {\em block algebra\/} that is 
associated to the chosen sequence of integrable weights, and denote it by
  \be \gb(\Pn) \equiv \gb\otimes_\complex \calf  \ee
with $\calf\equiv \calf({\dl P}^1{\setminus}\{p_1,p_2 \Ldots p_n\})$
($\calf(U)$ stands for the space of holomorphic functions on an open subset
$U\subset\tildeC$).

To arrive at the Ward identities, we have to define an action of the block 
algebra on $\vechil$, i.e.\ endow $\vechil$ with the structure of a 
$\gb(\Pn)$-module. The idea is to perform an expansion in local coordinates 
and to identify the local coordinates with the indeterminate of the loop space 
construction of the affine \lie\ $\g$. Accordingly, we introduce 
local coordinates $\xi_i$ such that $\xi_i(p_i)=0$, e.g.\ 
  \be \xi_i := z-z_i \ee
when the points $p_i$ correspond to values $z_i$ of the quasi-global coordinate
$z$ on ${\dl P}^1$. In terms of these coordinates we have local expansions
  \be  f(\xi_i) = \sum_{n\ge n_0} f_n^{(i)}\, \xi_i^n  \ee
of the functions $f\iN \calf$ (here the infinite sum
starts at some finite value $n_0$ which may be negative).
For every insertion point $p_i$ this local expansion induces a \lie\ 
homomorphism (actually, even an injection)
$\jmath_i$ from $\gb(\Pn)$ to the loop algebra $\g_{\rm loop}$.
Namely, the loop algebra is given by $\gb\otimeS\complex[[t,t^{-1}]]$ with
some indeterminate $t$ (thus $\g_{\rm loop}$ is essentially the affine algebra 
$\g$ without central extension), and $\jmath_i$ acts as
  \be \bar x\oT f\,\mapsto\, \jmath_i(\bar x\ot f):=\sum_{n\ge n_0} 
  f_n^{(i)}\, \bar x\oT t^n \,.  \labl{tn}
Actually, by making use of the residue theorem, one checks that this 
construction embeds the block algebra $\gb(\Pn)$ as a Lie subalgebra
into the $n$-fold direct sum $\g^n$ of the affine \lie; moreover,
one can check that this way the $\g^n$-module $\vechil$ indeed
acquires the structure of a module over $\gb(\Pn)$.

\subsubsection{WZW blocks}\label{swb}

In this framework, the space of \cb s can essentially be characterized as the 
space of singlets in the tensor product $\vechil$. Closer inspection shows 
that the relevant \rep\ theoretic concept is in fact the one of 
{\em co-invariants\/} $\coi{\vechil}{\gb(\Pn)}$ of $\vechil$ \wrt the 
block algebra \cite{tsuy,fesv2,beLa}.
The idea is to divide out the submodule that consists of all vectors in 
$\vechil$ that can be obtained by acting with an element of $\gb(\Pn)$ on
some other vector of $\vechil$; thus
  \be  \coi{\vechil}{\gb(\Pn)} = \vechil \,/\,\U^+(\gb(\Pn)) \vechil \,, \ee
where $\U^+(\gb(\Pn))\,{\equiv}\,\gb(\Pn)\U(\gb(\Pn))$ is the so-called 
augmentation ideal of the universal enveloping algebra $\U(\gb(\Pn))$ of the 
block algebra. 

For the benefit of those readers who are not familiar with this 
description of \cb s (that is frequently used in the mathematical literature), 
let us provide two pieces
of evidence that these are indeed the same objects that are encountered as
\cb s in the more familiar operator formalism. First, the co-invariants are 
{\em generalized singlets\/}. 
Namely, imagine that the tensor product $\vechil$ were fully 
reducible as a module over the block algebra $\gb(\Pn)$, i.e.\ that it can be 
decomposed into a direct sum of irreducible $\gb(\Pn)$-modules. 
Now in full generality, for irreducible modules $\hil_\lambda$
over any \lie\ $\gh$ we have $\U^+(\gh)\hil_\lambda=\hil_\lambda$ unless
$\lambda\eq 0$, in which case $\hil_0$ (the singlet) is \onedim\ while
$\U^+(\gh)\hil_0=0$, and hence $\coi{\hil_\lambda}\gh=0$ except for the singlet.
Thus in the situation at hand, quotienting out the subspace 
$\U^+(\gb(\Pn))\vechil$ of $\vechil$ would precisely leave us with the 
singlets in the tensor product space $\vechil$.
In other words, we would pick precisely those vectors in 
$\vechil$ that are invariant under all operators in the block algebra 
$\gb(\Pn)$; accordingly, this quotienting procedure indeed should correspond to 
implementing the invariance of the \cb s under the symmetries of the theory.

As a second hint, let us specialize to the particular case where $n\eq2$ with
$z_2\eq0$ and where $\bar x\eq J^a\iN\gb$ is the element of the horizontal 
sub\alg\ that corresponds to $J^a_0\iN\g$. 
Then $x:=z^m\ot J^a_{}\iN\gb(\Pn)$ acts on the elements of 
$\hil_{\Lambda_1}{\otimes}\hil_{\Lambda_2}$ as 
  \be \begin{array}{ll}  (z^m\ot J^a_{}) \oT \bfe + \bfe \oT(z^m\ot J^a_{}) \!\!
  &= ((\xi_1+z_1)^m\ot J^a_{}) \oT \bfe + \bfe \oT ((\xi_2)^m\ot J^a_{}) 
  \\{}\\[-.8em]
  &= \dsum_{j=0}^\infty \binom mj z_1^{m-p}\, J^a_p \oT \bfe + \bfe \oT J^a_m
  \,.  \end{array}  \labl{211}
The so obtained expression is nothing but the `modified coproduct' 
$\bigtriangleup_{z_1}(J^a_m)$ that implements \cite{fefk3} the fusion 
product of the sectors at the level of the chiral \alg. 
Thus the action of the block algebra as defined above generalizes the modified 
coproduct $\bigtriangleup_z$ to the situation with $n\,{>}\,2$ insertions.

As a matter of fact it is {\em not\/} true that the tensor product $\vechil$ is 
fully reducible as a $\gb(\Pn)$-module;
to obtain genuine singlets, one would have to start with the algebraic dual 
$(\vechil)^\star_{}$ of $\vechil$, which is a much larger space than 
$\vechil$. Also, concerning the fiber bundle description of blocks one must 
be aware of the fact that while the algebraic dual gives rise to the
trivial bundle ${(\vechil)}^\star_{}\otimeS\calm_{g,n}$, the singlets must
be determined fiberwise so that the resulting subbundle of \cb s is
generically a non-trivial bundle. Actually, this bundle is naturally defined 
over an extended moduli space $\calm_{g,n}^{\rm ext}$
that on top of the ordinary moduli (insertion points and moduli
of the curve) also includes the choice of local coordinates at the insertion 
points. We therefore consider the subbundle 
  \be  (\vechil)^\star_0 \,\to\, \calm_{g,n}^{\rm ext} \labl{totbun}
of singlets in the trivial bundle $(\vechil)^\star_{}\,{\times}\,\calm_{g,n}
^{\rm ext}\to \calm_{g,n}^{\rm ext}$. On the total space of the bundle 
\erf{totbun} we have a free and fiber-preserving action of 
$U^n$, where $U$ is the group of reparametrizations
of the local coordinate. When we evaluate the functions in \erf{totbun}
on the highest weight vectors and take the quotient with respect to the
action of $U^n$, we just end up with bundles that are isomorphic to the 
bundles $V_{\hat C_g}(\vec p,\vec\lambda)$ considered in subsection \ref{sbcb}.
The bundle \erf{totbun}, however, contains
more information, in particular about the descendants. Notice that, since the 
descendants transform non-trivially under changes of local coordinates, it is 
natural to keep track of local coordinates and to work over 
$\calm_{g,n}^{\rm ext}$. Actually, the singlets in the algebraic dual 
${(\vechil)}^\star_{}$ are closely related to the so-called {\em multi-reggeon 
vertex\/} in the case of free \cfts\ (for references as well as
the inclusion of background charges, see \cite{dpfhls}). In this formalism
one does not specify the sectors for the insertions; rather, one 
takes the full chiral space $\hil = \bigoplus_{\Lambda\in\Iset}\hil_\Lambda$
\erf{hil} of states, which is the direct sum of all (chiral) sectors. 
The $n$-reggeon vertex is then given by the space of co-invariants of the 
block algebra that acts on the $n$-fold tensor product of $\hil$ with itself. 
The \cb s are then obtained by evaluating these singlets (which are elements of
${(\vechil)}^\star_{}$, i.e.\ linear forms on $\vechil$)
on the highest weight states of specific sectors, which precisely
projects out the contribution of the relevant sector. (Notice that this 
formalism is frequently not formulated on the level of {\em chiral\/} \cft.)

\subsubsection{Block \alg s for general \cfts}

We close this subsection with the following side remark.
For a general conformal field theory, we postulate the existence
of an analogue of the block algebra, i.e.\ of an algebra that encodes 
the global realizations of all symmetries of the theory. 
Such a global algebra $\cala_{g,n}(m)$ should exist for every point $m$
in the moduli space $\calm_{g,n}$. Roughly speaking, the relation between this
global algebra and the chiral algebra should generalize 
the relation between the algebra of global functions 
(possibly subject to certain conditions at marked points) and local germs 
of functions. As a consequence, it should be possible to embed 
$\cala_{g,n}(m)$ in the direct sum of $n$ copies of the chiral algebra. The 
\cb s are then described in the same way as
for \wzwts, namely as co-invariants in the tensor product of $n$ sectors under
the action of $\cala_{g,n}(m)$. Concretely, it should be possible to obtain 
such a formulation of \cb s in terms of co-invariants for general \cfts\ by 
translating the analogous coproduct formula that generalizes the expression 
\erf{211} \cite{fefk3,gabe3} to the geometric description in terms of the 
functions $\calf$, \resp\ of vector and tensor fields. In the case of the 
Virasoro \alg, a first step towards such a formulation is implicit in 
\cite{scHl5}.
We also note that the type of structure we have just sketched also appears in 
several other contexts; e.g.\ it allows to exhibit \cite{kuta} a close analogy 
between chiral blocks of \wzwts\ and the theory of automorphic forms.

\subsection{Full \cft}

\subsubsection{The oriented cover $\tildeC$}

To study the relation between chiral and full \cft,
we now assume that some consistent collection of vector bundles (\resp\
sheaves) of \cb s is given. We first observe that while chiral \cft\ is a 
mathematically deep structure, it evidently does not provide the correlators 
of a physical field theory on the non-oriented manifold $C$. Specifically:
\nxt The chiral blocks are (generically) multi-valued functions of the 
     insertion points (they are sections in a non-trivial bundle)
     whereas correlation functions should be single-valued as functions of the
     insertion points (`locality'). Moreover, the correlation functions should
     also be essentially (i.e., up to the Weyl anomaly) functions of the moduli 
     of the curve.
\nxt The space of \cb s generically has dimension larger than one,
     whereas the physical correlators should be unique.
\nxt Moreover, we have artificially fixed an orientation, whereas
     both orientations should be completely equivalent. 

While the locality requirement for the dependence on the insertion points
is obvious, the corresponding statement for the moduli dependence deserves
a further comment. Indeed, as it turns out the correlation functions are 
{\em not\/} genuine
functions of the moduli of the curve. Let us illustrate this with the example
of zero-point correlators, i.e.\ partition functions $Z$. One way in which a 
moduli dependence shows up is the dependence of $Z$ on the representative of 
the world sheet metric $\gamma$ within a conformal equivalence class; namely,
  \be Z[\eE^f\gamma] = Z[\gamma]\cdot \eE^{c S_L(f,\gamma)} \,,  \labl{223}
where $c$ is the Virasoro central charge and the Liouville action $S_L$
is characterized by the `trace-anomaly'
  $\frac\partial{\partial f}\, S_L(f,\gamma)\eq\frac1{48\pi}\,\sqrt\gamma
  \,R_\gamma$.
As a consequence, the partition function $Z(m)$ is a section in a (real)
line bundle over $\calm_{g,0}$. In the case of genus $g\eq1$, it is related
to the bilinear expression 
  \be Z(\tau) = \sum_{\lambda,\mu\in\Iset} Z_{\lambda,\mu}^{}\,
  \chii_{\lambda}^{}(\tau)\, (\chii_\mu(\tau))^*_{}  \labl{tpf}
of characters 
by $Z(m)\eq Z(\tau) |\sigma|^2$, where $\sigma$ is a nowhere vanishing section
of a projective line bundle over $\calm_{1,0}$. For more details, we refer
to \cite{frsh}. Note that the Weyl anomaly in \erf{223} is proportional to the
value of the Virasoro central element; in critical string theory this vanishes 
when also the ghost sector is included; accordingly we will neglect this 
subtlety.

As it turns out, the three issues listed above
are intimately linked and are resolved
by one and the same construction. The starting point is the idea that 
in order to obtain results that do not depend on the orientation, at a first
stage we should keep track of both possible orientations; this leads us to
introduce the notion of the {\em oriented cover\/} $\tildeC$ of the manifold 
$C$. Let us describe the construction of $\tildeC$ directly for the general 
case; that is, we neither assume that $C$ is orientable and we allow for 
boundaries.  Thus we denote by $C$ a real \twodim\ manifold, which is 
possibly unorientable and can have boundaries. For the tangent space at each 
point $p$ in the interior of $C$ we have two different orientations; 
we construct another manifold from $C$ by taking over every such
point $p$ two points, one for each orientation. This way we obtain 
a two-sheeted cover $\tildeC$ of $C$. We stress that $\tildeC$ is not only
orient{\em able\/}, but is even naturally orient{\em ed\/}. 
The underlying manifold $C$ is orientable if and only if the two sheets of
$\tildeC$ are disconnected. When $C$ has a boundary, 
then $\tildeC$ is a {\em branched\/} cover of $C$.
$C$ can be obtained from $\tildeC$ as the quotient by the involution
$\I$ that exchanges the two sheets. The boundaries of $C$ are just the fixed
points of this involution $\I$. The involution reverses the orientation, hence
it is an anti-conformal map.\,%
\futnote{Note that the lift from $C$ to $\tildeC$ is defined on the level 
of complex structures. Concerning the metric structure, the following
statement can be made. For a metric $\gamma$ on $C$ to represent, upon
lifting to $\tildeC$, the complex structure on $\tildeC$, all boundary
components of $C$ must be geodesics in the metric $\gamma$.  For example, 
two half-spheres can be matched smoothly, whereas this is not possible
for two opposite `caps' on the sphere.}

As an illustration, take $\tildeC$ to be the Riemann sphere $S^2$, represented 
by the complex plane plus one point at infinity and endowed with
a (quasi-)global coordinate $z$. The involution 
  \be  \Id{:}\quad z\Mapsto 1/\baR z  \labl{Id}
then corresponds to $C$ being the disk, while for 
  \be  \Ic{:}\quad z\Mapsto-1/\baR z \labl{Ic}
the manifold $C$ is the real projective space ${\dl P}\reals^2$, also known 
as the crosscap. 

\subsubsection{The chiral theory on $\tildeC$}

We now restrict our attention again to the case where $C$ is orientable and
without boundaries. Then the cover is a non-connected curve that is the 
disjoint union
  \be  \tildeC = C_1 \,\dot\cup\, C_2  \ee
of two copies $C_{1,2}$ of the orientable curve $C$ which are endowed with 
the two opposite orientations; the involution $\I$ interchanges the 
components $C_1$ and $C_2$. Our approach to formulate the full \cft\ on $C$ in 
terms of chiral objects is to first construct the oriented cover $\tildeC$,
then establish a chiral \cft\ on $\tildeC$, and finally arrive at the full 
theory by getting rid of the orientation dependence. Concretely, this requires 
that for every $n \iN\zetpluso$
we have to lift an $n$-point situation on $C$ to $\tildeC$. This results in
a $2n$-point situation on $\tildeC$, where the $2n$ insertion points $p_i$ and 
$\Bar p_i$ with $i=1,2\Ldots n$ are related by
  \be  \Bar p_i = \I(p_i) \,,  \labl p
so that in particular they satisfy (say) $p_i\iN C_1$ and $\Bar p_i \iN C_2$
for all $i\eq1,2\Ldots n$.

To make contact to the more conventional formulation, we remark that
this effective doubling of the number of insertion points is usually described
by the statement that `in a \cft\ on a closed orientable surface
there are left- and right-movers.' It is worth mentioning that the terms
left-mover and right-mover have their origin in
the study of the spaces of solutions to classical field equations;
in contrast, here such structures arise without the need to require
that there exists a Lagrangian description of the theory.
Also note that in the case of the sphere $S^2$, the relation \erf p is 
precisely what one usually wants to express when one says that $z$ and 
$\Bar z$, previously regarded as two independent complex variables (the latter 
being conventionally denoted by $\bar z$), are to be considered as each others'
complex conjugates,
  \be  \Bar z= z^* \,. \ee

Note that superficially our description of closed \cft\ merely constitutes a 
minor modification of more conventional expositions. But still this innocent 
change of perspective allows us to explain the existence of left- and 
right-movers (which elsewhere are often introduced in a somewhat heuristic 
fashion) in a concise way via the connection with the fixing of the orientation.
To dispose of the dependence on the orientation we simply have to divide out 
the anti-conformal involution $\I$. The main benefit of our approach, however, 
will be that it allows us to treat open \cft\ to a large extend along precisely
the same lines.

The geometrical unoriented world sheet $C$ can be identified with the quotient
$\tildeC/\I$ of the oriented cover by the anti-conformal involution $\I$.
Correspondingly we regard the full \cft\ on $C$ as being obtained by lifting 
this quotienting procedure to the level of \cb s or, when thinking in terms of 
an operator formalism, of `fields'. Thus a field of the full 
\cft\ corresponds to two chiral fields on the oriented cover $\tildeC$.
For the chiral objects we can apply the theory developed previously. But
we have to take into account the relation \erf p, which means in particular
that in the full theory manipulations with insertion points, such as limiting 
processes, have to be taken in a correlated way.

\subsubsection{Bi-blocks}\label{sbib}

The next observation is that while the prescription to obtain $\tildeC$
from $C$ is unique at the geometrical level, typically there will be an
ambiguity on how to lift this prescription
to the field theoretical level. First of all, it need not
necessarily be required that one has one and the same chiral algebra on the two
sheets $C_1$ and $C_2$; taking different \alg s leads to so-called {\em 
heterotic\/} theories. But even if the chiral algebras on both sheets are
identical, there is a priori no reason to take the sector label $\Bar\Lambda$ 
on $C_2$ equal to the sector label $\Lambda$ on $C_1$, or in other words, 
we are still allowed to choose a non-trivial pairing between the sectors on the 
two sheets. Before we study in more detail the consistency requirements that 
we must impose on the pairings, we wish to present a few more comments on the
\cb s on the oriented cover.
 
The individual \cb s as described in formula \erf{cb0} are, generically, not 
single-valued. To arrive at a single-valued \corfu, we must look for a
specific horizontal section of the bundle 
$\calv\eq\calv_{\vec\Lambda,\vec{\Bar\Lambda}}$ of \cb s for the prescribed 
sequence of `external' sectors $\hil_{\Lambda_i}{\otimes}\hil_{\Bar\Lambda_i}$
for both connected components of the orientable cover. This means that the 
\corfu\ is to be obtained by forming a suitable linear combination of the blocks
with fixed external sectors $\vec\Lambda, \,\vec{\Bar\Lambda}$
and arbitrary allowed `internal' sectors $\vec\mu, \,\vec{\Bar\mu}$.
In short, the correlators of the full theory on $C$ 
are {\em linear\/} combinations of the \cb s on $\tildeC$. 
Now since in the case of closed \cft\ the oriented cover
$\tildeC$ is not connected, the Ward identities on the two components
of $\tildeC$ factorize. In the case of \wzwts\ this follows from the fact that 
  \be  \gb(C_1{\cup}C_2) = \gb(C_1) \oplus \gb(C_2) \,. \labl{oplus}
Namely, when $p_1,...\,,p_n\iN C_1$ 
are the insertion points corresponding to sectors 
$\hil_{\Lambda_1},\,\hil_{\Lambda_2},...\,,$\linebreak[0]$\hil_{\Lambda_n}$ 
and $\Bar p_1,...\,,\Bar p_n\iN C_2$ are the insertion points for
$\hil_{\Bar\Lambda_1},\,\hil_{\Bar\Lambda_2},...\,,
\hil_{\Bar\Lambda_n}$, \resp, then the relevant tensor product space is
$\vec\hil=\vechil\otimeS\vechilt$ with 
  \be  \vechil:=\hil_{\Lambda_1}\otimeS \hil_{\Lambda_2}\otimeS\cdots\otimeS
  \hil_{\Lambda_n} \qquad{\rm and }\qquad
  \vechilt:=\hil_{\Bar\Lambda_1}\otimeS \hil_{\Bar\Lambda_2}\otimeS
  \cdots\otimeS\hil_{\Bar\Lambda_n} \,,  \ee
so that as a consequence of \erf{oplus} also the co-invariants factorize,
  \be \coi{\vechil\otimeS\vechilt}{\gb(C_1\cup C_2{\setminus}\{p_1,...,
  \Bar p_n\})} 
  = \coi{\vechil}{\gb(C_1{\setminus}\{p_1,...,p_n\})} \otimes 
    \coi{\vechilt}{\gb(C_2{\setminus}\{\Bar p_1,...,\Bar p_n\})}\,.   \ee
In short, the chiral WZW blocks on $\tildeC$ have a factorized form,
so that it is appropriate to refer to them as {\em bi-blocks\/}. Such a 
factorized form of the blocks is expected for all other closed \cfts, too.

The correlators of the full theory are then linear combinations of these
bi-blocks. In the case of a non-heterotic theory, this looks of course
like {\em bilinear\/} combinations of blocks of one `chiral half',
which is the description used in the more conventional treatment.

Now recall that for chiral \cfts\ there exists an operator formalism (which is 
fully established at genus 0 and 1, while at higher genus it still has to be
worked out for general \cfts, see the remarks in subsubsection \ref{scb}).
In particular, the \cvo s \erf{pc} can be multiplied
and possess chiral operator product expansions \cite{fefk3}. It is usually
taken for granted that analogous expansions exist also for the full theory,
but we would like to stress that this is a rather non-trivial assumption even
once an operator formalism has been established at the chiral level.
But let us nevertheless assume for the moment that indeed we are given an 
operator formalism not only for the chiral, but also for the full theory.
Then the above result can be understood as follows.
In order that the \corfu s can be single-valued, it is necessary that
the physical fields of the full theory are linear combinations of
bi-chiral objects of the type 
  \be \CVO\Lambda\mu\nu(z) \oT \CVOB\Lambda\mu\nu(\Bar z) \,\in\,
  {\rm Hom}(\hil_\mu{,}\hil_\nu)[[z,z^{-1}]] \otimeS
  {\rm Hom}(\hil_{\Bar\mu}{,}\hil_{\Bar\nu})[[\Bar z,\Bar z^{-1}]]\,;  \labl{OB}
Thus the role of the (primary) \cvo s on $C$ is taken over by analogous
tensor product maps, and these are to be combined linearly.
In particular, to every field on $C$ one thereby associates a pair
$(\lambda,\Bar\lambda)\iN\Iset{\times}\Iset$ of sector labels
rather than a single label $\lambda\iN\Iset$.
 
Using the abbreviation \erf{phi} for primary
\cvo s, the factorization of bi-chiral blocks is then interpreted as
  \be  \hsp{-.8} \begin{array}l  \langle\, 
  \phI_{\Lambda_1}(z_1)\ot\phI_{\Bar\Lambda_1}(\Bar z_1)\,
  \phI_{\Lambda_2}(z_2)\ot\phI_{\Bar\Lambda_2}(\Bar z_2)\,
  \phI_{\Lambda_3}(z_3)\ot\phI_{\Bar\Lambda_3}(\Bar z_3)\,
  \phI_{\Lambda_4}(z_4)\ot\phI_{\Bar\Lambda_4}(\Bar z_4)\,
  \rangle^{(\mu,\Bar\mu)}  \\{}\\[-.7em] \hsp{4.8}
  = \langle
  \phI_{\Lambda_1}(z_1)\, \phI_{\Lambda_2}(z_2)\, \phI_{\Lambda_3}(z_3)\,
  \phI_{\Lambda_4}(z_4)\rangle^{(\mu)} \cdot
    \langle
  \phI_{\Bar\Lambda_1}(\Bar z_1)\, \phI_{\Bar\Lambda_2}(\Bar z_2)\,
  \phI_{\Bar\Lambda_3}(\Bar z_3)\, \phI_{\Bar\Lambda_4}(\Bar z_4)
  \rangle^{(\Bar\mu)} 
  \\{}\\[-.7em] \hsp{4.8}
  =: \calf_{\Lambda_1\Lambda_2\Lambda_3\Lambda_4}^{(\mu)}(z_1,z_2,z_3,z_4)\,
  \calf_{\Bar\Lambda_1\Bar\Lambda_2\Bar\Lambda_3\Bar\Lambda_4}^{(\Bar\mu)}
  (\Bar z_1,\Bar z_2,\Bar z_3,\Bar z_4)
  \end{array} \labl{oB}
for $n\eq4$
(here the labels $\mu$ and $\Bar\mu$ refer to the intermediate state,
compare the explanation of the corresponding notation in \erf{cb0}),
and analogously for general $n$.
Furthermore, comparison with the putative operator product expansion\,%
\futnote{The ellipsis stands for the contributions of descendant fields.
Also, any multiplicity labels that may be present are suppressed.} 
  \be   \begin{array}{l}
  \Phi_{\Lambda_1,\Bar\Lambda_1}(z_1,\Bar z_1)\, \Phi_{\Lambda_2,\Bar\Lambda_2}
  (z_2,\Bar z_2) \\{}\\[-.8em] \hsp{4.1} \sim \sum_{\Lambda_3,\Bar\Lambda_3} ^{}
  C_{\!\Lambda_1\Bar\Lambda_1,\Lambda_2\Bar\Lambda_2}^{\ \ \ \ \ \ \,
  \Lambda_3\Bar\Lambda_3}\, (z_1{-}z_2)^{\Delta_3-\Delta_1-\Delta_2}
  (\Bar z_1{-}\Bar z_2)^{\Bar\Delta_3-\Bar\Delta_1-\Bar\Delta_2}
  \, \Phi_{\Lambda_3,\Bar\Lambda_3}(z_2,\Bar z_2) + \ldots 
  \end{array} \labl{tO}
for the primary fields $\Phi_{\Lambda,\Bar\Lambda}$
of the full theory tells us that the coefficients that appear in the expansion 
  \be  \calg_{\Lambda_1\Lambda_2\Lambda_3\Lambda_4;
  \Bar\Lambda_1\Bar\Lambda_2\Bar\Lambda_3\Bar\Lambda_4}^{}
  = \sum_{\mu,\Bar\mu} \Conetofour\,
  \calf_{\Lambda_1\Lambda_2\Lambda_3\Lambda_4}^{(\mu)} \,
  \calf_{\Bar\Lambda_1\Bar\Lambda_2\Bar\Lambda_3\Bar\Lambda_4}^{(\Bar\mu)}
  \labl{FB}
of the \corfu s \wrt the (bi-)blocks \erf{oB} are nothing but
suitable products of operator product coefficients.

\subsubsection{Consistency conditions}\label{scond}

Concerning the \corfu s of the full theory, even independently of any operator
formalism it is clear that they will be suitable combinations of the \cb s.
More precisely, certainly not any arbitrary combination of blocks will
qualify as a sensible \corfu. Rather, various strong restrictions apply;
they are of the following three types:
\nxt \mbox{}{\em Locality\/}: 
     While the \cb s are sections in a (generically) non-trivial bundle
     over $\calm_{g,n}$, the correlators of the full theory must be 
     ordinary functions of the insertion points
     which provide (part of) the coordinates on $\calm_{g,n}$.
\nxt Analogously, we impose the requirement that the
     same statement applies to the dependence
     on the other coordinates on $\calm_{g,n}$, i.e.\
     on the moduli of the complex curves $C_i$.
     (The moduli that correspond to the two disconnected components of
     $\tildeC$ are to be identified via an anti-conformal involution, too.
     For instance, in the case where $C$ is the torus and hence 
     $\tildeC= C_\tau\cup C_{\Bar\tau}$, for the partition function 
     $Z(\tau)\eq\sum_{\lambda,\mu} Z_{\lambda,\mu}\chii_\lambda^{}(\tau)
     (\chii_\mu(\tau))^*_{}$ the relevant identification 
     is taken into account by the complex conjugation of $\chii_\mu$,
     which amounts to identifying $\Bar\tau$ with $-\tau^*$.)
\nxt \mbox{}{\em Factorization\/}: 
     The theory should be compatible with singular limits
     on the moduli spaces, in such a way that all coefficients
     that appear in the expansions of \corfu s in terms of \cb s
     (such as the $\Conetofour$ that appear in \erf{FB}) are
     expressible through the coefficients for the three point functions. 
     By comparing different sequences of factorizations that lead to one
     and the same final result, this requirement leads to various consistency
     relations, which are known as factorization or also as sewing constraints.
\\[.2em] 
   The presence of such locality and factorization constraints is in fact
   a necessary prerequisite for the existence of operator product expansions
   like \erf{tO}. In particular, the correct factorization of four-point
   correlators \erf{FB} amounts to the statement that the operator product
   \erf{tO} is associative.
\\[.2em] Finally there is also another type of constraints:
\nxt \mbox{}{\em Integrality\/}:
     The coefficients that appear in some specific
     linear combinations must be integral and non-negative. This applies to the 
     zero-point (bi-)blocks for surfaces with Euler characteristic zero
     (including those with boundaries or crosscaps).
     The reason is that the corresponding correlators of the full theory
     should acquire the physical meaning of partition functions,
     i.e.\ of generating functions for multiplicities of states of the full 
     theory.
\vskip.2em

It is believed that these constraints admit a unique solution for the
\corfu s; this has been checked in various \nontriv\ examples.
But note that the locality constraints alone can in general possess
several distinct solutions, corresponding e.g.\ to different possible
torus partition functions; for concrete realizations in the case of \sutwo\ 
\wzwts\ and $c\,{\le}\,1$ minimal models see e.g.\ \cite{fukl,dotr,petk2}.
 
We should also mention that the integrality constraints are in fact not
independent of the locality and factorization constraints.
In the case of closed \cft, where they apply to the torus partition function,
the crucial structure is provided by the fusion \alg, and 
integrality can be derived as a consequence of the Verlinde formula.
Similarly, in the open case the relevant concept turns out to be the one
of a {\em classifying \alg\/}, which is an associative \alg\
that generalizes the fusion \alg. This structure, introduced in \cite{fuSc5},
will be studied in subsection \ref{scla}.

Moreover, as intermediate `channels' that appear in a factorization formula\,%
-- e.g., as combinations $(\mu,\Bar\mu)$ in a factorization of the four-point 
function \erf{FB} into three-point functions -- only those combinations $(\mu,
\Bar\mu)$ of sector labels appear which are compatible with the chosen pairing 
$\om$, i.e.\ for which $\Bar\mu\eq\om(\mu)$. This requirement has an immediate
conceptual consequence: just like we already did at the level of chiral \cft, 
we can now also talk about considering one and the same full \cft\ on different
surfaces. Then the factorization constraints imply in particular that for 
each full \cft\ the pairing of labels has to be identical
on all closed oriented surfaces. In particular, we can then think of the 
pairing as being prescribed by the form
  \be  Z_{\lambda,\mu}\eq \delta_{\mu,\om(\lambda)} \equiv
  \delta_{\mu,\Bar\lambda} \,.  \labl{dlm}
for the torus partition function \erf{tpf}.

{}From now on we restrict our attention to torus partition functions of the
particular form \erf{dlm}. This is justified by the fact that as soon as
we talk about fusion rules and factorization at the level of the {\em full\/}
theory, we implicitly assume that no further extension of the chiral algebra
is possible that would alter the structure of the (chiral) fusion rules.

The factorization conditions severely constrain the possible
pairings $\om$. Namely, the various factorization limits must be compatible for
the chiral theories on both sheets simultaneously. On the other hand,
at the chiral level much of the information on the factorization is 
encoded in the fusion rules. As a consequence, the pairing $\om$ has
to be an automorphism of the fusion rules.\,%
\futnote{In the case of heterotic theories, analogously we need
an isomorphism between the fusion rules of the two chiral theories.}
Furthermore, the locality constraint requires that in addition the 
automorphism of the fusion rules does not change the conformal weight 
modulo integers of the primary fields.
This observation constitutes a natural origin for the appearance 
of different modular invariants and puts the general result \cite{mose2}
about the possible structure of modular invariants for theories with 
maximally extended chiral algebras in its natural context.

\subsubsection{Fusion rule \auto s}\label{sauto}

We have seen that the factorization constraint requires that the pairing 
$\om$ must be an {\em automorphism of the fusion rules\/}. This means that 
$\om$ satisfies
  \be  \N{\om(\lambda),}{\om(\mu)}{\,\ \ \ \om(\nu)}= \N\lambda\mu\nu 
  \qquad{\rm and}\qquad  \om(\vac)= \vac  \,.  \ee
Moreover, locality implies that $\om$ must commute with the action of the 
modular transformation $T$ that sends $\tau$ to $\tau{+}1$.
These fusion rule \auto s $\om$ of course form a group, with 
the unit element provided by the identity map on $\Iset$. But apart from this
property, the identity \auto\ does not seem to play any distinguished role,
in particular there is no reason to regard the corresponding {\em diagonal\/}
pairing
  \be  \Bar\Lambda = \Lambda  \labl{diag}
as more fundamental
than any non-trivial pairing. Indeed, it is typically not even
clear whether two full \cfts\ that are obtained through different choices 
of the pairing correspond to physically distinct situations.
The standard \nontriv\ example for a fusion rule \auto\ $\om$ of the required 
form is the {\em charge conjugation\/} \auto\ $\om_C$
which is given by the mapping \erf{mm}, i.e.
  \be  \Bar\Lambda = \om_C(\Lambda) \equiv \Lambda^+  \labl{cc}
for all $\Lambda\iN\Iset$. This \auto\ is present for every \cft\ (of course
it coincides with the identity map when all sectors are self-conjugate).
As far as we know, in all applications the charge conjugation cannot be
distinguished from the identity \auto\ by any physical property.

We also remark that in certain cases the fusion rule \auto\ $\om$
that is encountered 
in this context possesses a field theoretic realization. Namely, it can happen
that $\om$ is induced by the operation of forming the fusion 
product with some specific sector $\hil_J$. The relevant sectors $\hil_J$
constitute so-called {\em simple currents\/} \cite{scya,intr,scya6,jf28},
which in turn are closely related \cite{fusS3} to automorphisms of the chiral 
algebra. Via this connection fusion rule \auto s can correspond to \auto s of 
\chir, but clearly in the situation at hand it
is the \auto\ of the fusion rules that matters, independently of whether it
has a counterpart for the chiral \alg.\,%
\futnote{In this context one should note that
simple currents also play a role already in chiral \cft, where they give rise
to extensions of the chiral \alg\ \chir\ \cite{scya}. Moreover, {\em every\/}
simple current corresponds to an outer \auto\ of \chir\ which, however,
usually does not leave the Virasoro algebra invariant.
One might expect that such \auto s show up in open \cft\ as
well; for some comments on this issue see \cite{fuSc7}.\\
Note that even when an automorphism $\om$ of the fusion rules can be described
in terms of some simple current $J$, their actions on a general sector
need not coincide. For instance, the \sutwo\ \wzwt\ at level $k$
has a simple current $J$ which maps the sector with highest weight $\Lambda$ 
(which takes values in $\Iset\eq\{0,1\Ldots k\})$ to the sector $k\mi\Lambda$. 
In contrast, the automorphism of the fusion rules leaves sectors with 
$\Lambda\iN2\zet$ (i.e., integral isospin) invariant and only maps fields 
with odd $\Lambda$ to their simple current transform $k\mi\Lambda$.}
The identity \auto\ trivially possesses such a field theoretic realization,
the relevant simple current being just the vacuum sector. In contrast,
the charge conjugation \auto\ (when \nontriv) can never be interpreted in
this manner.

Finally, as an illustration consider the theory of a single free boson.
It is readily verified that in the uncompactified case, where the sectors are 
labelled by their charge $q\iN\reals$, the charge conjugation 
  \be  q\Mapsto-q  \labl{qq}
is in fact the only \nontriv\ fusion rule \auto\ that preserves the conformal 
weights. This remains true for a 
compactified boson, for which the (chiral) charges lie on a \onedim\ lattice.
When we have several, say $d$, free bosons, the situation is a bit more
complicated. In the uncompactified case, where the fusion product just 
amounts to addition of vectors $\vec q\iN\reals^d$, now every invertible
linear map of $\reals^d$ constitutes an automorphism of the fusion rules, i.e.\ 
instead of $\reals^\times_{}\,{\equiv}\,\GLe$ they now form the group $\GLd$.
Imposing the additional requirement that the \auto\ preserves the conformal 
weights $\Delta(\vec q)\eq \vec q^{\,2}/2$ of the sectors, this gets 
restricted to the orthogonal group \Od\ which generalizes $\zet_2\equiv\Oe$.
Finally we note that the group \Od\ contains in particular the element $-\one$.
The corresponding automorphism is just the charge conjugation \auto. 

\sect{Open \cft}

\subsection{The oriented cover}\label{soco} 

The presentation of our somewhat non-standard view of closed \cft\ took 
quite some time. We do think that this effort is rewarding for the study of
closed \cft\ itself. But it pays off even more once we turn to the study of
open \cft. Namely, we will now see
that once the appropriate formulation of closed \cft\ has been achieved, the 
extension to open \cft\ does not pose any major conceptual problems any more.
Indeed, to formulate \cft\ on a world sheet that has boundaries or
is unoriented, we follow exactly the same steps as in the closed case,
the main difference being that now the oriented cover $\tildeC$ of the 
unoriented surface $C$ is connected. In particular, we have again a map
  \be  \I: \quad \tildeC \to \tildeC  \labl{Io}
which is an anti-conformal involution. 

We have already presented this involution above for the examples of the 
crosscap and the disk. 
Similarly, for the annulus, the Klein bottle and the M\"obius strip the
oriented cover is a torus. More precisely, these three surfaces
are all characterized by a modular parameter $t\iN\reals_{\ge0}$, and
the annulus can be obtained from a torus with
modular parameter $\tau\eq\ii t/2$ by quotienting out $I_{\rm a}{:}\;
z\mapsto1{-}\baR z$, the Klein bottle from a torus with modular parameter 
$\tau\eq2\ii t$ via $I_{\rm k}{:}\; z\mapsto1{-}\baR z{+}\tau/2$, and
the M\"obius strip from a torus with modular parameter 
$\tau\eq(1{+}\ii t)/2$ via $I_{\rm m}{:}\; z\mapsto1{-}\baR z$.
For a more detailed exposition of surfaces with negative Euler characteristic
we refer to \cite{bisa2}.
  
We also note that the oriented surfaces that arise this way as oriented covers
of open or unorientable surfaces do not exhaust all possible 
complex curves of the appropriate genus; 
rather, their Teichm\"uller space can be embedded into the Teichm\"uller
space of the oriented cover. The latter Teichm\"uller spaces are actually
not only complex, but also symplectic manifolds; it is believed that the
Teichm\"uller spaces of the open surfaces form lagrangian submanifolds in
these spaces.  

\subsection{Bulk fields}\label{sbul}

Concerning the lift of the geometric prescription for going from $C$ to 
$\tildeC$ to the field theoretic level, as compared to the closed case two 
additional features have to be taken into account. In the language of the 
operator formalism, these features are expressed as follows.
First, to fields that are supported in the interior of $C$, which are called 
{\em bulk fields\/} \cite{card9}, one has to associate again a pair of \cvo s\,%
\futnote{The use of the oriented cover with two pre-images for points in
the bulk is reminiscent of the method of mirror charges that is employed to 
deal with other problems with boundaries conditions, e.g.\ in electrodynamics.}
on the cover $\tildeC$; the new aspect is that now one is no longer dealing 
with a tensor product map as in \erf{OB}, but rather one has to consider
a product 
  \be \CVO\Lambda\mu\nu(z) \odot \CVOB\Lambda\mu\nu(\Bar z) \,\in\,
  {\rm Hom}(\hil_\mu{,}\hil_\nu)[[z,z^{-1}]] \odot
  {\rm Hom}(\hil_{\Bar\mu}{,}\hil_{\Bar\nu})[[\Bar z,\Bar z^{-1}]]  \labl{OO}
that is formal in the sense that 
its precise meaning depends on the particular surface under consideration
as well as on the possible presence of further fields and has to be made
more concrete below. At first sight, this might
look like a rather big difference to the operator formalism of the closed
case, but in fact it is but another realization of the simple fact that the
oriented cover is now a connected manifold. 
The lesson to be drawn from this observation is then again that
we should better aim at a formulation of the chiral theory directly
in terms of the blocks. In fact, we will again try to employ the
concepts of block \alg s and co-invariants.
More precisely, the relevant modules will again be based on {\em ordinary\/}
tensor products of sectors $\hil_\Lambda^{}$ and $\hil_{\Bar\Lambda}$
with some suitable pairing of $\Lambda$ and $\Bar\Lambda$, but in
distinction from the closed case the block \alg\ will no longer have the
form of a direct sum of two subalgebras.

The second new feature arises when $C$ has boundaries, which happens when
the involution $\I$ \erf{Io} does not act freely any more. 
In this case there is an additional structure that was not present in the
case of closed \cft, namely the so-called {\em boundary fields\/} \cite{card9}
which live on the boundary of $C$ and accordingly correspond only to a single 
\cvo\ on $\tildeC$. These objects will be discussed in more detail later; for 
the moment, we concentrate on the case without boundaries where this 
complication is absent.

In the sequel we will study \cb s on various surfaces. Among them,
the one-point blocks are actually the most important ones, because more
complicated situations can be reduced to them with the help of factorization
arguments \cite{lewe3} that are completely analogous to those already employed
in the case of closed \cft. 
Indeed, invoking factorization, arbitrary $n$-point blocks on 
arbitrary surfaces can be expressed in terms of only three types of special 
building blocks: The three-point chiral blocks for the sphere (these are all 
that is needed in the case of closed \cft) as well as, as new ingredients 
needed for the open case, the one-point blocks (`tadpoles') 
on the sphere and the one-point blocks on the crosscap.

In the next subsections we will study the chiral \cft\ on $\tildeC$
in some detail. Afterwards, in subsection \ref{sful} we move on to the
full theory on $C$. Let us stress that in particular the one-point blocks
on the crosscap and on the disk that we present in subsections
\ref{scros} and \ref{sdisk} are {\em not\/} the physical one-point functions,
but rather the latter are to be obtained as suitable multiples of the former.

\subsection{Blocks on the crosscap}\label{scros}

The simplest surface without boundaries that has to be studied for open \cft\
is the {\em crosscap\/}, i.e.\  the real projective space ${\dl P}\reals^2$. 
(We speak about `the' crosscap because there is no modular parameter for this 
surface.)
The oriented cover of ${\dl P}\reals^2$ is the complex projective
space ${\dl P}^1$. Thus an $n$-point situation on the crosscap is mapped to 
a $2n$-point situation on ${\dl P}^1$. Writing ${\dl P}^1$ as the 
complex plane (compactified by a point at infinity) with coordinate $z$,
the relevant involution $\I$ is given by $\Ic$ \erf{Ic}, i.e.\
  \be  \Ic(z)=-\Frac1{\baR z} \,.  \ee
A fundamental domain for the action of $\I$ is given by the disk $|z|\,{\le}\,1$
with identification of diametrically opposite points of the circle $|z|\eq1$.
 
Our aim is to compute the one-point blocks for a bulk field with sector
labels $\Lambda$ and $\Bar\Lambda$. 
This situation corresponds to a two-point situation on ${\dl P}^1$,
with sectors $\hil_\Lambda^{}$ and $\hil_{\Bar\Lambda}$. 
In the spirit of our approach to \cb s, we want to construct this
\cb\ as a co-invariant of the tensor product space
$\hil_\Lambda^{}{\otimes}\hil_{\Bar\Lambda}$ \wrt 
an appropriate action of some block \alg.
But before doing so, let us first mention what we expect from a more
heuristic viewpoint. Namely, from the knowledge about two-point blocks 
on ${\dl P}^1$ in the case of closed \cft, one expects immediately that
the one-point block on the crosscap is non-zero if and only if 
  \be  \Bar\Lambda=\Lambda^+\,.  \labl{heu}
 
To investigate this issue in a more rigorous manner, let us first specialize to 
the case of
WZW models. In this case we can definitely apply the language of co-invariants 
to the study of the modules $\hil_\Lambda^{}\otimeS\hil_{\Bar\Lambda}$. For
simplicity let us assume in addition that the insertions are at $z_1=0$ and 
$z_2=\Ic(z_1)=\infty$. Moreover, let us for the moment restrict our attention
to trivial pairing $\Bar\Lambda \,{\equiv}\,\om(\Lambda)\eq\Lambda$ (much of our 
discussion will, however,
translate with only minor changes to the case of general pairing $\om$).
Then the block algebra consists of elements of the form 
$\bar x\ot f$ with $\bar x\iN\gb$ and $f$ a function with poles only at 0 and 
$\infty$, and hence is spanned by the elements $x:=\bar x\ot z^n$ with 
$\bar x\iN\gb$ and $n\iN\zet$. The local expansion of $x$ at $0$ is 
$\bar x \ot \xi_1^n$ with $\xi_1$ the local coordinate at 0.
To find the local expansion at $\infty$, we first realize that
the antiholomorphic local coordinate is $-\fraC1{\baR z}$, 
so the correct holomorphic coordinate is $\xi_2=-\fraC1z$ \cite{ales}; 
as a consequence, the local expansion of $x$ is $\bar x \oT (-1)^n\xi_2^{-n}$. 
Upon identifying the local coordinates $\xi_{1,2}$ with the indeterminate
$t$ of the loop construction (just as we did in \erf{tn}), we then
conclude that we must consider co-invariants
of $\hil_\Lambda^{}\otimeS\hil_{\Bar\Lambda}$ \wrt the action of
the Lie \alg\ that is spanned by
  \be  J^a_n \oT\bfe + (-1)^n\, \bfe \oT J^a_{-n}  \labl{jj}
with $n\iN\zet$ and $a=1,2\Ldots{\rm dim}\,\gb$.
(Note that this looks somewhat similar to the formula \erf{211} that we
encountered in the closed case.)
 
Moreover, again as in the closed case, the co-invariants are 
generalized singlets. In particular
if one had complete reducibility, one would be able to identify every vector in 
the space of co-invariants with some vector $\crosslk$ in the tensor product
$\hil_\Lambda^{}\otimeS\hil_{\Bar\Lambda}$. One would then write the defining 
condition for co-invariants as an equation for that vector $\crosslk$; 
introducing the short-hands 
  \be  J^a_n\oT\bfe=:J^a_n \qquad{\rm and}\qquad \bfe\oT J^a_n=: \Bar J^a_n\,,
  \ee
this would read
  \be  \llb J^a_n + (-1)^n \Bar J^a_{-n}\lrb \crosslk = 0  \labl{condc}
for all $a\eq1,2\Ldots{\rm dim}\,\gb$ and all $n\iN\zet$. In the literature 
the solution to these equations has been called the {\em crosscap state\/}, but 
since it is not a genuine state but rather a chiral block, we prefer the more 
accurate term crosscap one-point block, or for brevity, {\em crosscap block\/}.

Expressed in a bit more mathematical terms, the relation between the local 
expansions at the insertion points 0 and $\infty$ is given by the action
of an anti-\auto\ $\sigmac$ of the affine \lie\ $\g$ that acts as\,%
\futnote{Actually we should think of this as an
anti-automorphism of the nilpotent subalgebra $\g_{-}$ of the affine \lie\
$\g$. This extends to an anti-\auto\ of the whole affine \lie\ $\g$
provided that one also changes the sign of the central element.}
  \be  \sigmac:\quad J^a_m \,\mapsto\, (-1)^m\, J^a_{-m}  \,.  \labl{aau}
Via the affine Sugawara relation $L_m \propto \kappa_{ab}\sum_{n\in\zet} 
\normord{J^a_n J^b_{m-n}}$ this extends to an anti-\auto\ of the Virasoro
\alg\ that acts as
  \be  L_m \,\mapsto\, - (-1)^m\, L_{-m}  \,.  \labl;
(Also, for a general chiral {\em Lie\/} \alg\ \liechir, the analogous
formula will read $Y_m^i\mapsto(-1)^{m+\Delta_i-1}$\linebreak[0]$Y_{-m}^i$.)
It follows that the crosscap state has the property that it is preserved by
the Virasoro algebra, in the twisted sense that
  \be  \llb L_n - (-1)^m\,\Bar L_{-n} \lrb \crosslk = 0 \, . \labl:

A {\em formal\/} solution to the equation \erf{condc} can be given \cite{ishi} 
for every $\Lambda$ that in the chosen pairing $\om$ gets combined with 
its conjugate sector, i.e.\ for which
  \be  \om(\Lambda)=\Lambda^+\,;  \labl{31}
since our formulas were adapted to the case where the pairing is trivial, i.e.\
   $\om(\Lambda)\,{\equiv}\,\Bar\Lambda\eq\Lambda$,
this just means that we need $\Lambda\eq\Lambda^+$ and hence
reproduces the heuristic result \erf{heu}.
This formal solution is often called an `Ishibashi state'.
But this expression is {\em not\/} a state in the usual sense; it is not only
an {\em infinite sum\/} of basis elements of the tensor product vector space 
$\hil_\Lambda\otimeS\hil_{\Lambda^+}$, but, moreover, each of the terms in the 
sum has length one. As a consequence, the formal expression is not even 
contained in the completion of this tensor product vector space \wrt its 
standard scalar product. (Also, an interpretation as an eigendistribution to 
some operator with sensible conformal properties, which could serve as a 
conceptual explanation of the non-normalizability, does not seem to be 
possible. In the free boson case, a candidate for such an operator does 
exist, namely the Fubi\-ni\hy Ve\-ne\-zi\-ano field $X$ itself, but this 
is not a proper field of the \cft.)

We do not write down the explicit form of the crosscap one-point block for the 
WZW case, which can be found in the literature \cite{ishi,lewe3}. More 
interesting for the application to string theory is the case of an 
(uncompactified) free boson (recall that we may think of the corresponding \cft\
as the \wzwm\ based on $\hat{\liefont u}(1)$). Then the chiral
sectors are labelled by their charge $q\iN\reals$. Assuming again
that the relation between the two labels $q$ and $\Bar q$ that are
attached to a bulk field is provided by the diagonal 
pairing \erf{diag}, i.e.\ that $\Bar q\eq q$, there is only a single sector
that is paired with its conjugate, namely the
vacuum sector $\hil_0$. The formal solution for the one point-blocks 
takes in this case the form \cite{clny3} 
  \be  \crossk = \exp \llb -\sum_{n>0} \Frac{(-1)^n}n\, \alpha_{-n}
  \Bar\alpha_{-n}\lrb |0\rangle \ot |0\rangle \,, \labl{coo}
where the vacuum $|0\rangle$ is the highest weight state of $\hil_0$.
Apart from the fact that the two sets of oscillators appear in a `coupled'
manner, this just has the form of a coherent state
($\alpha_n\equiv J_n$ are the Fourier modes of $J\eq\ii\partial X$).

In the operator formalism, the presence of the non-trivial anti-\auto\ 
\erf{aau} translates into the inclusion of an additional operator $\cco$ which,
heuristically, `creates a crosscap' \cite{clny3} and is to be inserted
at $|z|\eq1$, so that for each bulk field there is one chiral part to the 
right and the other chiral part to the left of this operator. The meaning of 
the formal product $\odot$ that we introduced in the formula \erf{OO} can then 
be made concrete. Namely, for one-point blocks one arrives at the recipe that 
they are to be interpreted as (using the shorthand \erf{phi} for \cvo s)
$\langle\phI_\Lambda^{}(z)\, \cco\ \phI_{\Bar\Lambda}(\Ic(z)) \rangle$
(for $|z|\,{<}\,1$) -- regarded as ordinary \cb s on ${\dl P}^1$. For 
higher-point blocks the prescription is
\be \langle\,\phI_{\Lambda_1}^{}(z_1) \cdots \phI_{\Lambda_n}^{}(z_n)
   \,\cco\ \phI_{\Bar\Lambda_n}(\Ic(z_n)) \cdots \phI_{\Bar\Lambda_1}(\Ic(z_1))
   \,\rangle \,.  \labl{Gamma}
The operator $\cco$ is known as the {\em crosscap operator\/} and can be 
sensibly described only at the level of the full rather than the chiral 
theory.\,%
\futnote{The crosscap operator should implement the anti-\auto\ \erf{aau}.
Therefore it can presumably be regarded as a twisted
(anti-)intertwiner in the sense that $\cco J^a_m = \sigmac(J^a_m)\cco$.}
Its explicit form was displayed in
\cite{prss2}; it involves in particular coefficients $\Gamma_\mu$ which
are determined by a system of linear equations involving
fusing matrices and operator products (in the one-point case the presence of
$\Gamma_\mu$ only changes the over-all normalization of the block).
 
\subsection{\Ttype s}\label{sttyp}

\subsubsection{The choice of pairing}

At this point it is appropriate to point out that apart from the restriction
to diagonal pairing in the previous subsection -- which was in fact chosen 
merely for keeping the presentation as simple as possible --
in our discussion of open \cft\ so far we did not
bother to say anything about the way that the two labels $\lambda$
and $\Bar\lambda$ of a bulk field $\pho\lambda$ are related.
Clearly, the precise form of this relationship plays a crucial role
for the theory. For instance, when in the case of a \wzwt\ instead
of the diagonal pairing the charge conjugation pairing $\om_C$ \erf{cc} is 
chosen, then the condition \erf{condc} gets replaced by
  \be  \llb J^a_n - (-1)^n \Bar J^a_{-n}\lrb \crosslk = 0  \ee
which again possesses a solution precisely when \erf{31} holds, but now
that condition is satisfied for {\em every\/} $\lambda$, i.e.\ each bulk field
$\pho\lambda$ possesses a non-vanishing one-point block on the crosscap.

To study this issue more systematically, we compare to the analogous
situation in closed \cft. In that case the relation was described 
(see subsubsection \ref{sauto}) by a pairing $\om$, which for consistency 
with factorization and locality was required to constitute an
\auto\ of the fusion rules and to be compatible with $L_0$ modulo integers.
Surfaces with boundaries (or unorientable surfaces) generically possess
moduli spaces as well, which have
singular limits and therefore give rise to factorization constraints. 
They should take an analogous form as in the closed case.
Correspondingly, from now on we will assume that the pairing
  \be  \om:\quad \lambda \Mapsto \Bar\lambda  \ee
is again a fusion rule \auto. In short, in order to fully specify the theory 
on the oriented cover $\tildeC$ it is necessary to specify an \auto\ $\om$ of 
the fusion rules, which then leads in particular to a definite prescription 
for the block \alg s. Accordingly, for definiteness we will refer to the 
crosscap block -- and more generally to the open \cft\ in which it is 
computed -- to be of {\em\ttype\/} $\om$. The factorization constraints will 
in particular link the pairing for bulk fields on various surfaces. In a first 
step, however, we will consider a single surface in its own right.

Just as in the case of closed \cft, it is not clear whether every arbitrary 
fusion rule \auto\ can be used. Similarly to the closed case we will address
this issue on two levels. We first work at the level of \cb s,
and later we discuss the possible realization in an operator calculus. 
We assume that not only in the crosscap one-point situation, but
in full generality, for every allowed \ttype\ $\tya$ there exists a 
locally free sheaf (i.e., roughly, a vector bundle)
$\calv_\tya$ of \cb s\ analogous to the sheaf $\calv\equiv\calv_\tye$ of
blocks that arises for the diagonal pairing $\Bar\lambda\eq\lambda$.

\subsubsection{Implementation on \cb s}

In this section we consider again the system of vector bundles belonging to all
\cb s for any genus and any number of insertion points.
The fact that we deal with an automorphism of the fusion rules, which means
in particular that
  \be  \N{\om(\lambda),}{\om(\mu)}{\,\ \ \ \om(\nu)}= \N\lambda\mu\nu \,,  \ee
tells us that for all possible values of $g,n$ and $\lambda_1,\lambda_2 \Ldots
\lambda_n$ the rank of the vector bundle $\calv_{\lambda_1 \ldots \lambda_n}$
over the moduli space $\calm_{g,n}$ coincides with the rank of the vector bundle
$\calv_{\om(\lambda_1) \ldots \om(\lambda_n)}$ over the same moduli space.

To make meaningful statements, we have to postulate a bit more structure.
Namely, we require that the automorphism is {\em implementable\/} in the sense 
that for every value of the parameters there is an associated
isomorphism $\Om \equiv \Om_{\vec\lambda}$ between these vector bundles
$\calv_{\lambda_1 \ldots \lambda_n}$ and $\calv_{\om(\lambda_1) \ldots 
\om(\lambda_n)}$. It is not clear under what conditions an automorphism 
of the fusion rules is implementable.\,%
\futnote{On the other hand, it can even happen that a fusion rule \auto\
can be pulled back to the chiral \alg\ \chir. As we will see shortly, this
is the case for all automorphisms that are implementable in an operator
calculus. A special example is given by charge
conjugation for the free boson, which corresponds to $J_n\,{\mapsto\,}{-}J_n$.
In this context we remark that in the literature
other proposals for a classification of branes have been made
(for certain specific classes of \cfts) \cite{kaok,staN3} in which
the guiding principle is an \auto\ of the chiral algebra instead of a fusion
rule \auto. As we will see, an automorphism of the chiral algebra alone does
not provide enough information to specify a boundary condition. Accordingly
we do not expect that those ideas can be extended to generic \cfts.}
Moreover, there are automorphisms
of the fusion rules which can be implemented in several inequivalent ways;
we will encounter examples for such automorphisms later on when we discuss
the corresponding aspects in an operator calculus.

In the sequel, we will always restrict our attention to
the group of implementable automorphisms, which we denote by $G$,
counting of course inequivalent implementations
separately. This group contains several interesting subgroups. First, there is
the subgroup $G_T$ of $G$ that consists of elements that 
commute with the modular matrix $T$, i.e.\
$\Delta_{\om(\lambda)}\eq\Delta_\lambda \bmod \zet$.
As we have seen, this subgroup is
the one that is relevant in closed \cft. Second, we have the 
subgroup $G_0$ of $G_T$ that consists of all elements that preserve the 
conformal weight exactly, i.e.\ $\Delta_{\om(\lambda)}=\Delta_\lambda$, not 
just modulo integers. If we assume that the boundary conditions\,%
\futnote{This concept will be introduced in subsection \ref{sbound} and be
studied in great detail thereafter.}
are required to preserve the conformal symmetry exactly,
this subgroup $G_0$ is the relevant group for open \cft.

The group $G_0$ always contains the charge conjugation \auto\ as a central 
element, but otherwise it depends, of course, on the theory under consideration.
In some cases, like e.g.\ the free boson, it turns out to be a Lie group, which
can have different connected components. In that case, the choice of a 
connected component will look like a `topological' choice in a space-time
interpretation. In case the dimension of the Lie group $G_0$ is non-zero,
there are continuous moduli as well. {}From the \cft\ point of view, 
both types of moduli are on the same footing, while their geometrical 
interpretation might look quite different.

\subsubsection{Operator formalism}\label{sopf}

Let us now make again contact to the operator formalism and assume that the 
\cb s of the theory can be understood in terms of matrix elements of
chiral vertex operators. We call an automorphism {\em implementable}
at the operator level if there exists a family 
  \be \theta_\om^{(\lambda)}: \quad \hil_{\lambda} \to \hil_{\om(\lambda)} 
  \labl{xy}
of maps that is consistent with the \cb\ structure. In the case of \wzwts, where
we can describe \cb s by means of co-invariants, compatibility with the
\cb\ structure means that on all finite tensor products
  \be  \vechil = \calh_{\Lambda_1}^{} \otimeS \calh_{\Lambda_2}^{} \otimeS 
  \cdots \otimeS \calh_{\Lambda_n}^{} \ee
of irreducible highest weight $\g$-modules the map
  \be  \vectau := \bigotimes_{i=1}^n\, \tausigma^{(\Lambda_i)}  \labl{vectau}
factorizes to the co-invariants, or more precisely, to {\em all\/} 
co-invariants for {\em any} genus. Clearly, every automorphism that is
implementable at the operator level induces an implementable automorphism 
of the system of conformal blocks. In contrast, even if a theory admits an 
operator calculus, the converse statement is highly non-trivial. 
(As an aside, we remark that the analogous maps on modules that are induced
by simple currents $J_i$, \resp\ by the associated \auto s of the chiral
\alg\ \cite{fusS3}, can be implemented on the \cb s
only if the product of the simple currents is the identity,
$J_1{\star}J_2{\star}\cdots{\star}J_n\eq\bfe$ \cite{fuSc+}. This
illustrates once again that in the present context the basic structure is given
by \auto s of fusion rules, not by \auto s of the chiral \alg.)

In the language of chiral vertex operators, the implementability of the map 
\erf{xy} should amount to the statement that the diagram
  \be \begin{array}{ccccc}
  & \hil_\mu & \stackrel{\displaystyle \IA\lambda\mu\nu(v;z)}
  {\displaystyle \mbox{---------------------}\!\!\!\longrightarrow} & \hil_\nu
  \\{}\\[-.5em]
  \tausigma^{(\mu)}\!\!\!\!\!\! & \downarrow & & \downarrow & 
  \!\!\!\!\!\!\tausigma^{(\nu)}   \\{}\\[-.5em]
  & \hil_{\om(\mu)} & \raisebox{-1.5em}
  {$\stackrel{\displaystyle \mbox{---------------------}\!\!\!\longrightarrow}
  {\displaystyle \IA{\om(\lambda)}{\om(\mu)}{\om(\nu)}(\tausigma^{(\lambda)}
  v;z)}$} & \hil_{\om(\nu)} \end{array}\ee
commutes for every $v\iN \calh_\lambda$, which in turn will
imply that conformal blocks are mapped to conformal blocks.

Actually, it is not difficult to see that the knowledge of the \cb s for three 
insertions of the vacuum sector on ${\dl P}^1$ for all choices of insertion 
points and local coordinates around these points is equivalent to the complete
knowledge of the chiral algebra of the theory. (Notice that as a vector
space the vacuum sector $\hil_\vac$ is just the vector space underlying
the chiral algebra itself.) As a consequence, the fact that an implementable
automorphism $\tausigma$ preserves this vacuum three-point block implies that
  \be \tausigma : \quad \hil_\vac \to \hil_\vac \ee
even constitutes an automorphism of the chiral algebra.

We are now finally in a position to give an example for an automorphism of the
fusion rules that can be implemented in two inequivalent ways. To this
end we consider the identity automorphism of the fusion rules of 
a chiral \cft\ for which each sector is self-conjugate,
but in which some sectors have a negative Frobenius\hy Schur indicator. 
(Such sectors should be thought of as analogues
of symplectic (quasi-real) representations in the theory of \lie s. For more 
information about the Frobenius\hy Schur indicator see \cite{bant5}.) On
the modules belonging to these primary fields, the implementing map $\Om$
can be chosen to be either the identity or the natural involution that is 
provided by the symplectic form. Another class of examples is provided by 
\cfts\  with fixed points of simple current actions \cite{prss3,fuSc5};
on the fixed point labels the action of the \auto\ of the fusion rules is 
trivial, but again $\Om$ can be chosen to be non-trivial.

These examples also nicely illustrate that an automorphism of the chiral
algebra is not a sufficient datum to specify an \ttype. Namely, the associated
automorphism of the chiral algebra is the identity map in both
cases. However, the extension of the automorphism to sectors other
than the vacuum contains non-trivial information that cannot be reconstructed
from the automorphism of the chiral algebra alone. In particular, the
continuation of the automorphism to non-trivial sectors is not by simply
twisting the \rep\ $R_\Lambda$ of the chiral algebra on the sector $\hill$
by the automorphism, i.e.\ considering the action 
$R_{\om(\Lambda)}\,{\circ}\,\theta_\om^{(\Lambda)}$ on the same vector space 
that underlies $\hill$.  All these aspects are somewhat hidden in the case of 
the free boson, because in that case 
the only information about the algebra of chiral vertex operators that is
not already fixed by the chiral algebra can be encoded in the zero mode
of the Fubini\hy Veneziano field $X(z)$.

\subsubsection{Free bosons}\label{sfrb}

Let us now illustrate these points in the example of the free boson. In this
case we have a primary field for every $q\iN\reals$. Since the fusion rules
just realize charge conservation, $G$ is the group of all non-zero
real numbers $\alpha$,
where the group operation is multiplication; it acts as $q\To \alpha q$
with $\alpha\nE0$. Since the conformal weight is $\Delta_q\eq q^2/2$, the
subgroup $G_T$ contains just two elements corresponding to $\alpha\eq{\pm}1$. 
Thus there are two \ttype s; they turn out to correspond to
Dirichlet and Neumann boundary conditions for the free boson $X$.
The Neumann condition amounts to $\partial X\eq\Bar\partial X$, which 
means that momentum is conserved; in contrast, in the
Dirichlet case one deals with a brane that `carries momentum'.
Sometimes therefore the D-brane blocks are regarded as eigenstates of
the momentum operator and are therefore called `delocalized'
D-brane states \cite{dfpslr}. Heuristically, one would look for a
kind of Fourier transformation to find `D-brane states' that have a sharp
position. We will see below how this is afforded naturally in our formalism
on the level of full conformal field theory and how to choose the appropriate
linear combinations of D-brane blocks.

Next we consider the theory of $d$ free bosons. The primary fields
of the chiral \cft\ are labelled by their charges, which 
as already discussed at the end of subsubsection \ref{sauto},
now constitute a vector $q$ in $\reals^d$. The fusion product is addition 
in $\reals^d$, and the group of automorphisms of the fusion rules is 
$G\eq\GLd$. As a consequence, we can consider 
crosscap blocks $\crossk$, \resp\ the analogous objects for the disk,
the so-called boundary blocks $\boundk$ (see subsection \ref{sdisk} below),
that are invariant under the action of the relevant block algebra.
In the case of $\boundk$ this leads to the equation
  \be (\partial X^i - \sum_{j=1}^d M^i_{\,j}\, \Bar\partial X^j) \boundk = 0 
  \qquad\mbox{for all}\ i\eq1,2\Ldots d \, , \labl2
where $M\iN\GLd$. The subgroup of \auto s that preserve $L_0$ is $G_0\eq\Od$. 
One can check that in geometric terms the choice of $M$ corresponds to 
choosing the dimension of a D-brane and a field strength on the brane.

In the case of free bosons, chiral vertex operators can be given explicitly
using the string coordinates $X^i$ and their derivatives. Now of course 
operators living on disconnected sheets of the oriented cover commute. In the 
case of open surfaces, however, the oriented cover is connected, and hence
all chiral vertex operators must be concatenated. This is sometimes expressed 
by saying that in the closed orientable case (i.e.\ for bi-blocks)
there are two independent sets $\alpha_n\equiv J_n$ and
$\Bar\alpha_n\equiv\Bar J_n$ of oscillators,
while for all other surfaces there is just one set of oscillators.

The action of the charge conjugation pairing on the \cvo s is given by
$\eE^{\ii q X(z)} \mapsto \eE^{-\ii q X(z)}$ for a single free boson. For $d$ 
free bosons, the general automorphism is
  \be  \eE^{\ii q\cdot X(z)} \Mapsto \eE^{\ii (\CX q)\cdot X(z)} \,,   \labl{c3}
where $\CX \iN\Od$. The eigenspaces of $\CX$ to the eigenvalue $+1$
give the directions in which one has Neumann boundary conditions. Eigenvalues
$-1$ correspond to Dirichlet boundary conditions, while the rest corresponds to 
a field strength on the world volume of the brane. As an example, the matrix
${\rm diag}((+1)^{p+1}{,}(-1)^{d-p-1})\iN\Od$ corresponds to a Dirichlet 
$p$-brane  for which the field strength on the $p{+}1$-dimensional world
volume of the brane vanishes. 

It is common to rewrite
  \be  (\CX q)\cdot X(z) = q \cdot (\CTX X(z))  \ee
so that the set of all transformations \erf{c3} (for all relevant values of $q$)
just reduces to a single map
  \be  X(z) \Mapsto \CX X(z)  \,.  \labl{c4}
In the case of a general \cft\ it will no longer be possible to encode the
fusion rule \auto\ in such a simple formula, because no analogue of the quantity
$X(z)$ (which is not a genuine conformal field) is available any longer.

\subsubsection{Compatibility}

Up to this point, we have only discussed the pairing $\om$ on a single given
surface. We now address the additional aspects that must be taken into
account when one requires consistent factorization
of all blocks on arbitrary surfaces.
Just like in the case of closed \cft, only after imposing this constraint
we can sensibly talk about the `same' \cft\ simultaneously on all surfaces.

In this context an important observation is that given a definite torus 
partition function,
non-trivial solutions for the one-point blocks will exist only for a subset of
the bulk fields, and it will depend on the choice of the \ttype\ what this
subset looks like. In particular, not any \ttype\ will allow for solutions
for sectors other than the vacuum sector.
More precisely, once we  have fixed both the torus partition function and
the \auto\ $\om$, we can study for which values of
$\lambda$ the tensor product $\hil_\lambda\otimeS\hil_{\Bar\lambda}$
possesses any co-invariants for the relevant block \alg. In the
case of the crosscap with trivial \ttype, for \wzwts\ this is just the \lie\
spanned by the modes \erf{jj}, and we can immediately conclude that there 
exists at least one $\lambda$ for which there is a (formal) solution,
namely $\lambda\eq\vac$, simply because the combination $(\vac,\vac)$ is
contained in every sensible torus partition function and every \auto\ of the
fusion rules leaves the vacuum invariant, $\om(\vac)\eq\vac$. But apart from
this special case the existence of solutions is not guaranteed. 
(Similarly, for generic $M\iN\Od$ the equation \erf2 will not have solutions 
other than $q\eq\Bar q\eq 0$.) On the other hand, it turns out that
when the torus partition function is given by charge conjugation and
the \auto\ $\om$ is the charge conjugation \auto\ as well, then
a solution exists for every $\lambda$.

As a side remark, we note that the element $-\one$ of $\Od$ allows to
flip between (generalized) Dirichlet and (generalized) Neumann boundary
conditions because it changes the relative sign of left movers and right movers.
Now recall from subsubsection \ref{scond} that we can restrict our attention
to theories for which the torus partition function is given by a fusion
rule \auto\ $\pi$. We will sometimes abuse terminology and employ the
terms {\em Neumann\/} and {\em Dirichlet\/} \ttype\ to refer to the 
situation where $\pi$ preserves conformal weights and where the pairing 
$\om$ conincides with $\pi$ and with its `charge conjugate' $\pi{\circ}\om_C$,
\resp.

Finally we note that the precise form of the co-invariants will of course 
depend on the chosen \ttype, too.
Consider for instance again the case of a single free boson. Denoting the
highest weight state of $\hilq$ by $|q\rangle$, for Neumann \ttype,
i.e.\ $\Bar q\eq q$, the crosscap one-point blocks read
  \be  \crossqk = \exp \llb -\sum_{n>0} \Frac{(-1)^n}n\, \alpha_{-n}
  \Bar\alpha_{-n}\lrb |q\rangle \ot |{-}q\rangle  \labl{don}
(which of course includes \erf{coo} as a special case), while for
Dirichlet \ttype, i.e.\ $\Bar q\eq q^+ \,{\equiv}\,{-}q$, they are
  \be  \crosskd = \exp \llb +\sum_{n>0} \Frac{(-1)^n}n\, \alpha_{-n}
  \Bar\alpha_{-n}\lrb |q\rangle \ot |q\rangle \,. \labl{doo}
Moreover, according to the previous remarks in the case of the diagonal
torus partition function the Neumann block \erf{don} exists only for $q\eq0$
while the Dirichlet block \erf{doo} exists for arbitrary charge $q\iN\reals$,
and in the case of the charge conjugation torus partition function 
(which is just the T-dual of the diagonal partition function that is usually 
chosen to describe the uncompactified free boson) the situation is reversed.

\subsubsection{Compactified free bosons}\label{scfb}

To illustrate these considerations, we consider again a system of $d$ free
bosons, this time compactified on some $d$-dimensional torus $T^d$. On this
torus, we choose a basis $\{e_i^{}\}$ and denote by $\{e_i^\star\}$ 
a dual basis. Such a compactification is characterized by the metric 
$g_{ij} := (e_i,e_j)$ and an antisymmetric tensor $B_{ij}$. The torus partition 
function relates the left and right moving charges $q_{L,R}^{}$ according to
  \be  q_{L,R}^{} = \sum_{i=1}^d \llb m_i^{} e_i^\star \pm \Frac12\, 
  n_i^{} e_i^{} \lrb
  - \Frac12 \sum_{i,j=1}^d B_{ij}\, e_i^\star n_j^{} \,,  \ee
where $n_i$ and $m_i$ are integers. The vectors $q_L{\oplus}q_R$ form a
self-dual sublattice $\Gamma(g,B)$ of $\reals^{n,n}$ (i.e., 
of $\reals^{2n}$ with signature $(n,n)$). In string theory terms, $m_i$ is 
a momentum number and $n_i$ a winding number; the background field $B$ 
couples to the winding.

We now study a fixed \ttype, described by some orthogonal matrix $M$.
To find out which bulk fields will lead to boundary blocks, we consider the
$d$-dimensional subspace
  \be \cald_M := \{ (q,Mq) \}  \ee
of $\reals^{2d}$. For
a chosen background $g,B$, the set of those pairs of left and right charges
for which there are non-vanishing one-point blocks for the \ttype\ specified
by $M$ is given by the sublattice
  \be \Gamma(g,B,M) := \Gamma(g,B) \cap \cald_M  \labl{gbm}
of $\Gamma(g,B)$.
For fixed $g,B$ the rank of $\Gamma(g,B,M)$ depends on $M$; generically it
is zero. For example, for Neumann \ttype, i.e.\ $M\eq\one$, only for discrete
values of the Kalb\hy Ramond background field $B$ non-trivial
one-point blocks exist at all.\,%
\futnote{This quantization of $B$ plays is crucial in the description
of type I duals of the CHL string \cite{bian}.}

We conclude this subsection with a brief comparison to the geometric description
of branes. At the level of chiral \cft, only aspects of the topology of the
brane (in particular its dimensionality) and the field strength on it enter.
Other features like e.g.\ the value of the moduli of the world volume of the
brane will only show up on the level of full \cft, see subsubsection \ref{sEx} 
below.

\subsection{Boundary conditions and boundary fields}\label{sbound}

Having clarified the structure of chiral \cft\ for the case of the crosscap,
we now proceed to study what happens when we allow for the surface $C$
to have boundaries. To start, we still restrict our attention to bulk
fields. As pointed out in subsection \ref{soco}, due to factorization
the quantities of prime interest
are the one-point correlation functions of the bulk fields.

Now when imposing the various consistency conditions detailed in subsubsection
\ref{scond}, notably the integrality constraint 
for the partition function on the annulus, there will be severe restrictions on
the possible one-point functions. As it turns out, these restrictions 
typically allow not only for one, but for several solutions. 
(This is in sharp contradistinction to the closed case, where the locality
and factorization constraints are believed to possess a unique solution.) To 
every such solution, i.e.\ to every consistent collection of one-point 
correlation functions for all bulk fields, one commonly associates a 
corresponding {\em \bc\/}.

Except for the case of free fields, where one just deals with \bc s
in the usual geometric sense, it is a priori not clear how to 
translate the notion of \bc\ that is introduced this way directly to
a prescription on how the fields behave when they come close to the boundary.
In fact, this is an issue that concerns the full rather than the chiral \cft,
and accordingly will be studied in more detail in subsection \ref{sful} below.
But already at this point we can say that in the operator formalism
the appropriate characterization of a \bc\ is \cite{card9} as a collection 
of certain {\em reflection coefficients\/}, whose conceptual status is 
similar to the one of operator product coefficients.
   
Roughly, one can imagine that the boundary carries some kind of
`charge'. The bulk fields `feel' this charge when they approach the boundary.
But even without invoking an operator formalism for the full nor even for the
chiral theory, we can
already be more specific at the level of the chiral blocks. Namely, just
like in the case of the crosscap, we have the freedom to choose an \ttype\
that fixes the pairing of the bulk labels, and certainly the blocks will depend
on the chosen \ttype. Thus the \bc\ includes in particular a definite
prescription for the \ttype\ that is to be chosen, i.e.\ is to some extent
specified by a label $\tya$. In addition, however, the \bc\ may contain 
further information, corresponding to some other label for which we generically
use the symbol $a$. Accordingly, we should think
of a \bc\ as a {\em pair\/} of labels $\om$ and $a$.\,%
\futnote{Similar ideas have been expressed in the recent paper \cite{reSC}.}

Since the presence of the label $a$ is tied to the very
existence of boundaries, one may refer to the corresponding freedom as the
{\em boundary type\/}. On the other hand, in order to be close to
conventional terminology, we continue to refer to the pair consisting of both
$\om$ and $a$ as a \bc, and accordingly the term boundary type may be slightly
misleading. Therefore we prefer to use a different term, namely {\em\ctype\/};
this is inspired by the role that these quantities will play in string theory,
where to every pair $\Oma$ one can associate an independent so-called 
\cp multiplicity (see subsection \ref{scp} below).
Of course, when we restrict our attention to ordinary Neumann \ttype\
(which is the situation usually considered in the \cft\ literature, e.g.\
in \cite{card9,lewe3}), we may wish to suppress the label for the \ttype\
and speak of the \bc\ and the corresponding \ctype\ $a$ interchangeably.

Note that whereas we have a quite clear prescription that tells us what the
various possible \ttype s are, so far we have been somewhat loose concerning
the possible values of the \ctype\ $a$. Indeed we do not want to study
this issue in any detail here; rather, this will be done in subsection
\ref{scla}, yielding the concept \cite{fuSc5} of a {\em classifying \alg\/}.

Next we turn our attention to boundary fields, which we commonly denote by
$\phb{}(x)$. The distinguished feature of boundary fields is that they
`live' on the boundary of the surface $C$ \cite{card9}, i.e.\ $x$ is a 
coordinate on $\partial C$. Upon lifting to the oriented cover $\tildeC$,
boundary fields should therefore correspond to a single \cvo, and accordingly 
they carry a single sector label $\Lambda$. (In string theory terms, where 
the bulk fields -- more precisely, Virasoro-primary bulk 
fields of conformal dimension 1 -- correspond to the vertex operators for closed
strings, boundary fields correspond to the vertex operators for open strings.\,%
\futnote{Recall that in the free boson case the primary \cvo s are nothing 
but the ordinary vertex operators 
$\normord{\eE^{\ii\lambda\cdot X(z)}}$.} \mbox{$\!\!$}\,)
In addition, however, since they are confined to the boundary, the
boundary fields will possess an
intrinsic dependence on the \bc\ that is attached to the connected component of
$\partial C$ on which they are inserted, and accordingly they will also
carry a corresponding label $\oma\,{\equiv}\,\Oma$.
More precisely, in fact there must be {\em two\/} such labels. The reason is 
that the boundary fields also play the role \cite{card9} of effecting a
change of boundary conditions, say from $\oma$ to $\omb$. (Such changes of 
boundary conditions indeed occur not only in string theory, but also
in various situations that are of interest in condensed matter physics, see 
e.g.\ \cite{affl8} and the literature cited there.)
Thus we should describe boundary fields as
  \be  \phb{}(y) = \pHB\oma\omb\Lambda(y)  \ee
when we have \bc s $\oma$ for $x\,{<}\,y$ and $\omb$ for $x\,{>}\,y$, \resp.

Note that an immediate consequence of this description is that
besides the boundary fields that change only the \ctype\ $a$
which have been studied e.g.\ in \cite{card9,lewe3,prss3}, there must in fact
also exist boundary operators that change the \ttype.
(Incidentally, such `topology changing' operators are also
implicit in \cite{kont7}.)
Geometrically, this corresponds to the situation that a boundary of the world 
sheet jumps from one brane to the other, at a point where the world volumes of 
the branes intersect.

Finally we note that we have not been too specific about the label $\Lambda$
that is attached to boundary fields. Below we will usually assume that 
$\Lambda$ already appears as a sector label of the bulk theory, i.e.\ that
$\Lambda\iN\Iset$. But in principle more general choices can lead to consistent
theories as well.  Roughly speaking, if we use a chiral \cft\ $\calc$ to 
construct \cb s on surfaces without boundaries (including unorientable 
surfaces), we can choose for a given \ttype\ $\om$ a \cft\ $\tilde C$ with a 
`bigger' fusion algebra to construct \cb s in the presence of boundaries. The
chiral theory $\tilde\calc$ cannot be chosen independently from $\calc$,
though. Namely, consider in the space $\hil$ of chiral states of $\calc$ 
the subspace $\hilo$ of those states which are allowed as bulk fields for the
\ttype\ $\om$. We then require that one can embed $\hilo$ into
$\tilde\hil$ and that $\tilde\calc$ has some implementable automorphism 
$\tilde\om$ of the fusion rules that prolongs $\om$ under the embedding. In 
short, one imposes an additional projection in the bulk as compared to the 
boundary.

As an example, we consider again the free boson compactified on a circle of
radius $R$. For $\tilde\calc$ we choose the free boson
compactified on a circle of radius $\ell R$, were $\ell$ is a natural number. 
The Dirichlet blocks of $\tilde\calc$ are given by $\boundqk_{\rm D}^{}$, 
defined as in \erf{boo} below, with $q\eq n/\ell R$ (for more details, 
see subsubsection \ref{sEx}). The embedding of the D-allowed bulk 
fields of $\calc$ into the D-allowed bulk fields 
of $\tilde\calc$ is obvious, and these are all obvious embeddings of D-allowed 
sectors of a free boson into D-allowed sectors of another free boson.
We want to argue that this situation corresponds to a brane that wraps $\ell$ 
times around the circle. First, we observe that the boundary conditions in a
standard theory based on $\tilde\calc$ are given by
$\tilde a\bmod 2\pi\ell R$.
As a consequence, the range for the theory $\tilde\calc$ wraps $\ell$ times
around the range of the theory $\calc$. However, the bulk cannot probe these
details: its momenta are only $q\iN \zet/R$. As a consequence,
  \be \eE^{\ii q (\tilde a + 2\pi R)} = \eE^{\ii q \tilde a} \,, \ee
and since we probe the localization with bulk fields (we use left
and right movers), $\tilde a$ and $\tilde a{+}2\pi R$ are localized
at the same point. But still, these are different boundary conditions:
we have different 3-point functions for three insertions on the boundary,
and on the boundary fields with different conformal weights appear.

\subsection{Blocks on the disk}\label{sdisk}

We now briefly discuss chiral one-point blocks for bulk fields on the disk.
Like for the crosscap, the oriented cover of the disk is ${\dl P}^1$, which we
write again as the complex plane plus a point at infinity, so that
the relevant anti-conformal involution is given by $\Id$ \erf{Id}, i.e.\
  \be  \Id(z)=\Frac1{\baR z} \,.  \labl{Idisk}
Just as for the crosscap there is no modular parameter for the disk.\,%
\futnote{Thus all disks with arbitrary values of the radius are conformally
equivalent. For definiteness here the disk is always taken to have unit radius.
To describe a disk of arbitrary radius $R$, one simply would have to
replace the map \erf{Idisk} by $z\mapsto R^2/z$, or equivalently, rescale the
local coordinates around both pre-images of an insertion point by $R$.}

We first restrict our attention to Neumann \ttype. Then the analysis of the
one-point blocks parallels the analogous derivation for the crosscap in
subsection \ref{scros}.
The only difference is that in place of the anti-\auto\ $\sigmac$ we have
to use a corresponding anti-\auto\ $\sigmad$.
This differs from $\sigmac$ just in the omission of the factor $(-1)^m$
that is a sign of the non-orientability of ${\dl P}\reals^2$. Thus e.g.\
in the case of \wzwts\ it reads
  \be  \sigmad:\quad J^a_m \,\mapsto\, J^a_{-m}  \labl{smd}
instead of \erf{aau}, for the Virasoro \alg\ one has the corresponding
analogue of \erf;, and similarly for other chiral \alg s.

Just like in the case of the crosscap, the one-point block is constructed as
a co-invariant of the tensor product $\hil_\Lambda^{}\otimeS\hil_{\Bar\Lambda}$
\wrt the action of the block \alg. And just like in that case it is common
to ignore the fact that this tensor product isn't fully reducible as a module
over the block \alg, and correspondingly use the notation $\boundlk$ for the
one-point block and write the defining condition of the co-invariants as
  \be  (J^a_n + \Bar J^a_{-n}) \boundlk = 0 = (L_n - \Bar L_{-n}) \boundlk  
  \labl{jl}
analogous to \erf{condc} and \erf:.
Clearly, a formal solution to these conditions is obtained from the one for 
$\crosslk$ by simply removing the appropriate factors of $(-1)^n$. For
instance in the case of a single free boson, instead of \erf {coo} one now has
  \be  \boundk = \exp \llb -\sum_{n>0} \Frac1n\, \alpha_{-n} \Bar\alpha_{-n}
  \lrb |0\rangle \ot |0\rangle \,,  \labl{bnk}
while the analogues of the more general crosscap blocks \erf{don} and \erf{doo} 
read
  \be  \boundqk = \exp \llb -\sum_{n>0} \Frac1n\, \alpha_{-n}
  \Bar\alpha_{-n}\lrb |q\rangle \ot |{-}q\rangle  \labl{bon}
and
  \be  \braneqk \equiv \boundqk_{\rm D}^{} = \exp \llb +\sum_{n>0} \Frac1n\,
  \alpha_{-n} \Bar\alpha_{-n}\lrb |q\rangle \ot |q\rangle \,, \labl{boo}
\resp.

In the literature this formal solution is known as a boundary state (or also 
Ishibashi state); but we stress again that it has the status of a {\em chiral
block\/} and accordingly prefer to call $\boundk$ a {\em boundary block\/}.

The boundary block \erf{bnk} is the one-point block for the vacuum bulk field.
Depending on the chosen pairing, the conditions \erf{jl} may or may not
possess a (formal) solution also for other bulk fields. Moreover, one can
perform an analogous analysis also for other \ttype s than the Neumann
type, in which case there is a different (action of the) block \alg, 
leading to generalized boundary blocks $\boundk_\om$. In the special
case of Dirichlet \ttype, and also more generally when they correspond to
brane configurations in string theory, these are often denoted by $\branek$
and referred to as {\em brane states\/}. These issues
can again be treated completely parallel to the corresponding discussion in the
crosscap case, and we will not repeat this here.
What {\em is\/} different to the crosscap case is that for a given collection
of {\em one-point blocks\/} (with fixed \ttype) there typically exist
several distinct consistent collections of {\em one-point correlators\/},
namely precisely one for each \ctype;
this will be studied in more detail in the next subsection.

\subsection{Correlators of the full theory}\label{sful}

For the very same reasons as in the case of closed \cft, in the open case
the correlation functions of the full \cft\ on $C$ are again to be
constructed as specific linear combinations of \cb s on $\tildeC$.
And again they are severely constrained by locality, factorization and
integrality constraints. Since the oriented cover is now connected,
the blocks can no longer be written in the form of bi-blocks, though.

In particular, the one-point correlators of bulk fields
$\pho\Lambda$ on the disk and on the crosscap
are linear combinations of the relevant boundary and crosscap blocks, \resp.
Now in this particular case according to the results of the previous
subsections there exists only (at most) a single chiral block, and hence these
one-point correlators are simply proportional to the one-point blocks.
In the case of the crosscap with Neumann \ttype, the constant of
proportionality is nothing but the coefficient $\Gamma_\Lambda$ that appears
(compare the remark after \erf{Gamma}) in the crosscap operator \cite{prss3}.
In contrast, in the case of the disk the
proportionality constant depends on the \bc\ $\oma\,{\equiv}\,\Oma$.
More precisely, up to the $\Lambda$-independent factor
\futnote{For the presence of this factor and the explanation why this 
normalization constant is generically different from unity we refer to 
\cite{prss3,fuSc7}.}
  \be  \Alpha^{AA}_\vac := \langle\, \pHB AA\vac \,\rangle  \labl{av}
it is given by the {\em reflection coefficients\/} $\rc A\Lambda\vac$
of the bulk field \wrt the vacuum boundary field; thus the one-point
correlation function of $\pho\Lambda$ on the disk with \bc\ $A$ reads
  \be  \langle\pho\Lambda\rangle_{\!A}^{} = \Alpha^{AA}_\vac\,\rc A\Lambda\vac\,
  \boundlk_\om \,.  \labl{pha}

The reflection coefficients that show up here are usually introduced (and 
receive their name) in the operator formalism, namely by the expansion 
\cite{lewe3,prss3}
   \be  \pho\Lambda(r\eE^{\ii\sigma}) \,\sim\, \sum_{\mu\in\Iset}\sum_{a\in\Io}
   (r^2{-}1)^{-2\Delta_\Lambda+\Delta_\mu}_{}\, \rc A\Lambda\mu\,
   \pHB AA\mu(\eE^{\ii\sigma}) \qquad {\rm for}\;\ r\to 1  \labl{pp}
in terms of boundary fields, which the bulk field should possess when 
$|z|\sim1$. Thus they encode how the bulk field
behaves close to a boundary component, or in other words, to what extent it
`excites' the boundary fields when it approaches the boundary.
Note that in \erf{pp} it is assumed that the \ttype\ $\om$ is fixed
(in particular $\Bar\Lambda\equiv\om(\Lambda)$ with $\om$ the prescribed 
\ttype); the second summation in \erf{pp} is then over the set $\Io$ of
all those \ctype s $a$ for which there is a \bc\ $A\eq\Oma$.
Moreover, to be precise, the expansion \erf{pp} was established in
\cite{lewe3,prss3} only for the case of \bc s of Neumann \ttype; here we 
assume that it remains true for any other \ttype, 
which, employing the existence of an operator formalism, follows \cite{reSC} 
by the same arguments as in the Neumann case. Similar remarks apply
to all other relations that follow in this subsection; an argument why this 
assumption should be valid will be presented in the next subsection.

The formula \erf{pha} nicely displays the physical meaning of the \bc s
$A$. Namely, they constitute a degeneracy index $A$ that labels the
various consistent collections of one-point correlators on the disk.
Note that the allowed linear combinations are
severely restricted by the locality and factorization requirements,
and in particular by the integrality and positivity constraints for
the annulus, M\"obius strip and Klein bottle; these have been discussed in
great detail in \cite{lewe3,prss3}.
Now whereas it is expected that the linear combinations of bi-blocks that 
constitute the correlators of a closed \cft\
are uniquely determined by the various constraints,
as it turns out, in the presence of boundaries there typically indeed
exist several different consistent solutions, i.e.\ several distinct
\bc s, even for prescribed \ttype\ $\om$. 
In general, it is a rather difficult task to obtain a
classification of all possible \bc s by studying the various constraints.
As we will propose in the next subsection, there exists, however, a general
structure which neatly encodes the set of all \bc s, namely the so-called
classifying \alg.

Another quantity in which the reflection coefficients $\rb A\Lambda\vac$
that appear in \erf{pha} enter is the zero-point \corfu\ on the annulus, which
is briefly called the annulus amplitude. To investigate this
quantity, we consider the combinations
  \be  \bA:= \sum_{\Lambda\in\Iset} \langle \pho\Lambda \rangle_A^{}
  =\sum_{\Lambda\in\Iset} \Alpha^{AA}_\vac\, \rb A\Lambda\vac \, \boundlk_\om^{}
\,. \labl{bA}
Here and in the sequel we shall adopt the convention that the boundary blocks
are normalized in such a way that 
  \be  {}_\om^{}\boundlb \, q^{L_0+\tilde L_0 -c/12}_{}\,
  \boundlp_\om^{}
  = \delta_{\Lambda,\Lambda'}^{}\, (\Frac{S_{\vac\Lambda}}{S_{\vac\vac}})^{-1}\,
  \chii_\Lambda(\tau) \, , \ee
where $\chii_\Lambda(\tau)$ is the Virasoro-specialized character. Note that
in the case of free conformal field theories, the prefactor $S_{\vac\Lambda}/
S_{\vac\vac}$, the so-called quantum dimension, is equal to one.
In the literature, the particular linear combinations \erf{bA}
of boundary blocks are again often referred to as boundary states.

These boundary states do not possess any immediate physical meaning. They 
should rather be compared to the $n$-reggeon vertex that was mentioned
at the end of subsubsection \ref{swb}, for which one also considers expressions 
that are direct sums over all sectors of the space of chiral states. Indeed, 
saturating one leg of the $n$-reggeon vertex with a boundary state $\bA$ 
amounts to introducing a boundary of type $A$ on the world sheet. Also in this 
case the physical correlators are to be obtained by applying suitable 
projections to the boundary state. For example, the
classical $p$-brane solutions of the type II superstring can be
recovered this way \cite{dfpslr}.
Using factorization, we can express the annulus amplitude as the `product'
of two such boundary states and one `propagator':
  \be  \begin{array}{ll} A^{AB}(t) \!\!&
  = \langle {\cal B}^A |\, \eE^{-\frac{2\pi}t(L_0+\Bar L_0-c/12)}_{}\,
  |{\cal B}^B \rangle \\{}\\[-.8em] &
  = \dsum_{\mu\in\Iset} \Llb \Frac{S_{\vac\mu}}{S_{\vac\vac}} \Lrb^{-1}_{}
  (B^A_\mu \Alpha^{AA}_\vac)^*_{}\, \chii_\mu(\Frac{2\ii}t) \,
  (B^B_\mu \Alpha^{BB}_\vac) 
  = \dsum_{\mu\in\Iset} A_\mu^{AB}\,\chii_\mu(\Frac{\ii t}2) \,.  
  \end{array} \labl{aa}
Here the second expression and the last one
are related by a modular transformation. In the string context,
\erf{aa} is the partition function for the open string states (before 
orientifold projection), and the second expression corresponds to the
closed string channel while the last one describes the open string channel.
This short calculation also nicely illustrates that due to factorization 
arguments we can obtain a complete overview over 
the boundary conditions from considerations on the disk alone.

To conclude this subsection, let us mention a few other aspects of the
operator formalism \cite{lewe3,prss3}. First there is an operator product 
expansion for two boundary fields; it reads 
\futnote{Here $\Delta^{\rm b}_\lambda$, which according to \erf x 
governs the decay of a boundary field two-point function along the boundary, 
need not coincide with the ordinary (bulk) conformal dimension 
$\Delta_\lambda$ of the sector $\hil_\lambda$ \cite{card9}.}
  \be  \phb\lambda^{AB}(x_1)\, \phb\mu^{BC}(x_2)  \sim
  \sum_{\nu\in\Iset} (x_1\mi x_2)^{\Delta^{\rm b}_\nu-\Delta^{\rm b}_\lambda-
  \Delta^{\rm b}_\mu}_{} C^{ABC}_{\lambda\mu\nu}
  \phb\nu^{AC}(x_2) \qquad{\rm for}\;\ x_1\to x_2\,.  \ee
Combining this formula with the requirement that the only boundary
fields with non-vanishing one-point function are the vacuum fields
$\pHB AA\vac$ and with the relation \erf{av}, 
it follows that the two-point blocks of boundary fields read
  \be  \langle \phB{AB}\lambda(x_1)\, \phB{BA}\mu(x_2)\rangle
  = (x_1-x_2)_{}^{-2\Delta^{\rm b}_\lambda} \Alpha_\lambda^{AB}\,
  \delta_{\lambda^{},\mu^+} \,,  \labl x
with
  \be  \Alpha_\lambda^{AB} = C^{ABA}_{\lambda\lambda^+_{\phantom I}\vac} \cdot
  \Alpha_\vac^{AA} \,.  \ee

Finally we consider the sewing constraint that comes from the correlation
function $\langle \pho\lambda(z_1)\, \pho\mu(z_2)\,\phb\nu^{AA}(x_3) \rangle$ 
for two bulk fields and one boundary field. This correlator
can be factorized either by first using the operator product
of two bulk fields and afterwards the reflection coefficient for the resulting
bulk field, or else using twice the reflection coefficients. According to 
\cite{lewe3,prss3} this provides in particular the relation
  \be  C_{\!\lambda\Bar\lambda,\mu\Bar\mu}^{\ \ \ \ \nu\Bar\nu}
  \rc A\nu\vac \Alpha_\vac^{AA}
  = \sum_{\kappa\in\Iset} \Alpha_\kappa^{AA} \epsilon_{\lambda\mu}^\nu
  \eE^{\ii\pi (\Delta_\mu-\Bar\Delta_\mu+\Delta_\kappa)}
  \rc A\lambda\kappa \rc A\mu\kappa
  \Fmat\kappa\nu{\lambda^{}}{\mu^{}}{\mu^+}{\lambda^+} \,,  \labl y
which besides the coefficients introduced above contains the entries of
fusing matrices $F$ and certain sign factors $\epsilon$ (the latter are
related to the Frobenius\hy Schur \cite{bant5} indicator).

\subsection{Classifying algebras}\label{scla}

\subsubsection{Boundary conditions as \rep s of an algebra}

Our goal is now to make more definite statements about the possible boundary 
conditions. To this end we implement the information from the conventional 
operator formalism for the full \cft\ \cite{card9,lewe3,prss3} that we
collected in the previous subsection. But while we need this as a heuristic 
input, we expect that our conclusions in fact do not rely on the existence of 
an operator formalism at all and can be replaced by statements about \cb s and 
the structure of their singularities when a curve degenerates.

Specifically, we take the formula \erf y and contract it by the inverse fusing 
matrix $\Fmatm\nu\vac{\lambda^{}}{\mu^{}}{\mu^+}{\lambda^+}$. This amounts to 
\cite{prss3}
  \be  \rc A\lambda\vac\, \rc A\mu\vac
  = \sum_{\nu\in\Iset} \Nt\lambda\mu\nu\, \rc A\nu\vac \labl{class}
with some numbers $\Nt\lambda\mu\nu$ which are combinations of of operator 
product coefficients, entries of fusing matrices and normalization constants.

The equation \erf{class} proves to be an extremely useful result. In contrast to
the crosscap constraint \cite{prss2} which is linear, it provides us with 
non-linear relations and therefore also allows to fix the normalization of the 
reflection coefficients.\,%
\futnote{Actually, the fact that three crosscap insertions
are topologically equivalent to one crosscap and a handle provides an 
additional constraint. Namely, the crosscap coefficients $\Gamma_\Lambda$ must 
obey a relation in which they enter both cubically and linearly. This 
constraint should also allow to fix their absolute normalization.}
First recall that \erf y has been obtained in the literature \cite{prss3} under 
the (implicit) assumption of dealing with \bc s of Neumann \ttype. But as
already indicated above, we expect that it holds in fact for any other
\ttype\ as well. As will be discussed next, the relation \erf{class}, 
gives rise to the structure of an associative \alg\ that looks
similar to the fusion \alg\ which governs the behaviour of ordinary
(Neumann) blocks. We regard this as an indication that the vector bundles
of blocks for arbitrary \ttype\ carry structures analogous to
those of the Neumann blocks, such as a \KZ\ connection and fusing and braiding
properties.\,%
\futnote{We also expect that just like in the case of Neumann \ttype,
the analogue of the \KZ\ connection is projectively flat 
and unitary. Moreover, as already pointed out above,
there should be factorization rules that relate different vector bundles over
different moduli spaces. 
The blocks should then still furnish a \rep\ of the modular group;  
it would be interesting to see whether one can describe the rank of these 
vector bundles of blocks through a generalization of the Verlinde formula.}
Given these structures, the derivation of \erf{class} will
be completely parallel to the Neumann case.
Further evidence comes from the study of examples, see below.

Anyhow, we take it for granted that the relation \erf{class} holds for 
every \ttype, and moreover, that it holds
independently of the existence of an operator formalism. 
We interpret the formula \erf{class} as follows. It tells us that
once an \ttype\ $\om$ has been prescribed, then for each fixed 
boundary condition $A\,{\equiv}\,\Oma$ the reflection coefficients 
$\rc A\Lambda\vac$ furnish a \onedim\ \irrep\ of an associative
algebra with structure constants $\Nt\lambda\mu\nu$.\,%
\futnote{Sometimes a different normalization convention for the reflection 
coefficients $\rb A\Lambda\vac$ is chosen and also the factor $\Alpha^{AA}_\vac$
\erf{av} is included. The so obtained reflection coefficients
have the serious disadvantage that they do not furnish a \rep\ of
a classifying algebra any more.} 
We call this \alg\ the {\em classifying algebra\/} for the boundary conditions
of type $(\tya,a)$ with fixed \ttype\ $\tya$ and denote it by $\CA_\tya$.
The determination of all possible boundary conditions, i.e.\ of all possible
values of $a$ for given \ttype\ $\tya$, is thereby reduced to the study of the 
representation theory of the \findim\ algebra $\CA_\tya$. 

\subsubsection{Properties of $\CA_\om$}

A distinguished basis of the classifying algebra $\CA_\om$
is labelled by the collection of bulk fields $\pho\Lambda$ 
that for the chosen \ttype\ $\om$ have a non-vanishing one-point function 
on the disk (that is, a non-vanishing boundary block $\boundlk_\om$).
We denote the corresponding basis elements by $\tom_\Lambda$, so that
the relations of the \alg\ read
  \be  \tom_\lambda\,\tom_\mu = \sum_{\nu\in\Iset} \Nt\lambda\mu\nu\, \tom_\nu
  \,.  \ee
Then the statement that the reflection coefficients furnish a \rep\ $\rp_a$
of $\CA_\om$ simply means that
  \be  \rc A\Lambda\vac = \rp_a(\tom_\Lambda)  \labl{rp}
for all allowed labels $\Lambda$.

Owing to factorization arguments, the classifying algebra $\CA_\om$ can be
expected to be an associative \alg\ over $\complex$.
Further properties of $\CA_\om$ follow from our general picture above
which relates the classifying algebra to the properties 
of \cb s with \ttype\ $\om$. In particular, analogously as in the case of the
fusion rule \alg\ we expect that $\CA_\om$ is commutative and that
the specific generator $\tom_\vac$ that is associated to the vacuum sector
is a unit element (in this regard, it is a nice consistency check that $\vac$ 
is an allowed label for {\em every\/} \ttype\ -- 
this would no longer be true if we would deal with arbitrary
automorphisms of the chiral 
algebra in place of fusion rule \auto s). Moreover, the evaluation at the 
unit element should provide a conjugation (involutive automorphism)
of the \alg, which in turn together with the other properties implies 
that the algebra is semi-simple. When the theory is rational, $\CA_\om$ is 
\findim, and therefore commutativity and semi-simplicity imply that it has 
only \onedim\ \irrep s, as many inequivalent ones as its dimension. 

Taking these properties for granted, without loss of generality we should 
also be allowed to require that the (equivalence classes of) irreducible 
\rep s of $\CA_\om$ are already exhausted by those \rep s which according to 
\erf{rp} are provided by the various possible \ctype s $a$.
In other words, the allowed \bc s for given \ttype\ are {\em precisely\/}
the \onedim\ \irrep s of the classifying \alg, so that in particular for a
rational theory their
number (which by other methods would be quite difficult to determine) is
  \be  |\Io| \equiv
  |\{A\eq(\om,a)\,{\mid\,}\om\;{\rm fixed}\}| = {\rm dim}\,\CA_\om \,.  \ee
In short, up to the information about 
the explicit form of the boundary blocks,
a boundary (or D-brane) state is nothing but a mnemonic for some
definite irreducible \rep\ of the classifying \alg.

A less obvious property of the classifying \alg\ concerns the
integrality and positivity 
of the structure constants $\Nt\lambda\mu\nu$ of the classifying algebra. In the
case of the fusion rule \alg, the structure constants are non-negative integers
because they count the dimensions of spaces of \cb s. This needs no longer
be true for non-Neumann \ttype s $\om$. In the general case we would rather 
expect that, at least when $\om$ is implementable in the sense of subsubsection
\ref{sopf}, then in place of dimensions the structure constants will correspond
to traces on the spaces of \cb s. This generalizes the structure
of the classifying algebra that was found in \cite{fuSc5}. We expect that these
traces on the spaces of \cb s are related to twisted traces
in the sectors $\hill$ (or in other words, generalized character valued 
indices) similar to those that were studied in \cite{fusS3}, namely to traces of
operators that involve also the maps $\theta_\om^{(\lambda)}$ \erf{xy}. The 
relation between the traces on the sectors and the traces on the spaces of \cb s
should generalize the Verlinde formula which relates the modular
transformation properties of the ordinary characters to the dimensions of
the spaces of ordinary \cb s (which are special examples of traces, namely
of the unit matrix). In the context of open \cft, twisted traces have been 
considered in \cite{reSC} (compare equation (4.28) of \cite{reSC}). When 
the implementing maps \erf{xy} have order
two (as is e.g.\ the case for the classifying \alg\ that was obtained in
\cite{fuSc5}), then the structure constants $\Nt\lambda\mu\nu$ will
still be integers, though they are allowed to be negative; but for general 
order they even need not be integers any more.

As a final comment we point out that the description of boundary conditions
in terms of representations of some algebraic object is actually in nice
correspondence to the geometric description, say for type IIB compactifications
of the superstring on a Calabi\hy Yau manifold $\calm$. 
(For some background about boundary conditions in type II superstring
theories see \cite{oooy}.) In this case, specifying Dirichlet boundary 
conditions amounts to the specification of a coherent sheaf on $\calm$. 
Coherent sheaves, however, are just certain modules of the structure sheaf
of the manifold. Now the structure sheaf of $\calm$, i.e.\ the sheaf of local 
germs of holomorphic functions on $\calm$, can be thought of as a precursor of 
the chiral algebra `before quantization', 
so that also in the geometric approach the possible boundary conditions are
determined by the representation theory of an appropriate algebraic object.

\subsubsection{Examples}\label{sEx}

Unfortunately, though computable in principle, the quantities from
which one can calculate the structure constants $\Nt\lambda\mu\nu$ in
an operator framework, such as the fusing and braiding matrices and the
operator product coefficients, are so 
far not available for a generic \cft. They have only been worked out for 
\sutwo\ \wzwts\ (both with the diagonal and with non-diagonal modular 
invariants) and for Virasoro minimal models. 
It has been conjectured by Cardy \cite{card9} that (expressed with 
the help of the notions introduced above) the classifying algebra
for a rational \cft\ with charge conjugation modular invariant and boundary
conditions of Neumann \ttype\ just coincides with the fusion algebra 
of the chiral theory. The only case where this proposal has been verified
in an explicit calculation is for \sutwo\ \wzwts\ \cite{prss3,sasT2}, in which 
case as just mentioned the relevant data of the chiral \cft\ are fully known.

But even when these data are not known explicitly, our concept of classifying 
\alg\ turns out to be very fruitful. 
For instance, Cardy's conjecture follows as an immediate consequence
of our general picture, because for blocks of Neumann \ttype\ --
that is, in particular, for ordinary blocks in the case when the
torus partition function is given by charge conjugation --
the classifying \alg\ is nothing but the fusion rule \alg. Furthermore, 
by the same token, we also get the result that the classifying \alg\ 
for blocks of charge conjugation \ttype\ is the fusion rule \alg, too,
as soon as the torus partition function is the diagonal one.

As a check on the ideas presented above, we discuss a few more examples.
We start with the theory of a single uncompactified free boson. Then
the chiral sectors \hilq\ are labelled by the real numbers, $q\iN\reals$,
and the fusion product just realizes charge conservation, $q_1\star q_2= 
q_1{+}q_2$. In the torus partition function every diagonal combination 
$(q,q)$ of sectors appears precisely once. There are two \ttype s
which we denote by N and D; they correspond to Neumann \resp\ Dirichlet
boundary conditions for the free boson $X$. As seen in subsection \ref{sdisk}
there is only a single N-boundary block $\boundk\,{\equiv}\,\boundok$, and as 
a consequence only a single N-boundary condition. In contrast, there is a 
D-boundary block $\braneqk$ for every $q\iN\reals$; 
accordingly, the `D-brane states' are of the form
  \be  \dalpha = \int_{-\infty}^{\infty}\! \rmd q\, \rb aq\vac \braneqk 
  \, . \ee
According to our discussion above, the classifying algebra should be the 
fusion algebra. Hence we have to realize the relations $\RB a{q_1}{\Bar q_1}\vac
\RB a{q_2}{\Bar q_2}\vac\eq \RB a{q_1+q_2}{\,\Bar q_1+\Bar q_2}\vac$;
their solutions read
  \be  \rb a q\vac=\eE^{-\ii a q} \qquad{\rm with }\quad
  a\iN \reals \,.  \labl{aR}
As we will see below, the label $a\iN\reals$ that characterizes the \bc\
can be interpreted as the position of the D-brane. By comparison with the 
remarks at the end of subsubsection \ref{scfb}, this means that
we have recovered the last missing geometrical datum of the D-brane.

The free boson compactified on a circle of radius $R$ can be studied
in a similar way. The choice of a compactification radius is equivalent 
to a choice of a torus partition function for the boson; infinite radius 
(i.e., the uncompactified case)
corresponds to the diagonal partition function, and multiplication of the
partition function with the charge conjugation matrix amounts to going from the
radius $R$ to the T-dual compactification with radius $2/R$.
In all cases the fusion product, which is a chiral concept, is the same as for
the diagonal partition function, i.e.\ just expresses charge conservation.
At radius $R$, the fields that occur in the torus partition function have
charges $(q_L^{},q_R^{})=(n/R\,{+}\,2mR,n/R\,{-}\,2mR)$,
where the momentum number $n$ and the winding number $m$
take their values in the integers. It follows that there are infinitely many
boundary blocks of Neumann type, since only the momentum number is required to
vanish, $n\eq 0$, whereas the winding is arbitrary. For the case of Dirichlet
boundary conditions the situation is reversed; the winding number must
vanish, $m\eq0$, but the brane can carry arbitrary momentum $n$. 
In short, the Neumann blocks are $|{\rm B}_{2mR}\rangle$ with $m\iN\zet$
and the Dirichlet blocks are $|{\rm D}_{n/R}\rangle$ with $n\iN\zet$,
where $\boundqk$ and $\braneqk$ are as defined in \erf{bon} and \erf{boo},
\resp. This is compatible, of course, with T-duality which interchanges momentum
and winding states as well as Dirichlet and Neumann boundary conditions.
    
This time the classifying algebra is just the restriction of the fusion rule
\alg\ (which is the group \alg\ of $\zet{\times}\zet$) to the allowed 
sectors, i.e.\ both in the Neumann and in the Dirichlet case $\CA$ is 
the group algebra of $\zet$. We then obtain the formula
  \be  \rB an\vac =\eE^{- \ii an / R}  \ee
for the reflection coefficients; since $n$ takes its values in the integers, $a$
can now be restricted to lie in $\reals \bmod 2\pi R \zet$.
The interpretation of $a$ (and 
$n$) depends on whether we deal with Dirichlet or Neumann boundary conditions; 
in the case of Neumann boundary conditions $a$ is interpreted to come from a
U(1) background gauge field, while in the case of Dirichlet boundary
conditions $a$ is identified with the position of the D-brane.
This follows \cite{gree8,grgu} from the relation
  \be  \eE^{\ii q X} \,\dalpha
  \equiv \eE^{\ii q X} \dsum_{n\in\zet} \eE^{-\ii an/R} |{\rm D}_{n/R}\rangle
  = \dsum_{n\in\zet} \eE^{-\ii an/R} |{\rm D}_{q+n/R}\rangle
  = \eE^{-\ii a q}\, \dalpha \,.  \labl{333}
Notice that it is consistent to give $q$ the dimension of a momentum, i.e.\
the inverse of a length. Equation \erf{333} only makes sense if the momentum
$q$ takes its values in $\zet/R$. This is precisely the quantization
of momentum on a circle of radius $R$.

When the square of the compactification radius is a rational number,
$R^2\eq2r/s$ with $r$ and $s$ coprime, the free boson theory possesses further
symmetries so that it becomes a rational \cft, with $2rs$ sectors. 
One can then impose the additional requirement that the boundary
conditions preserve also these new symmetries. We can show that this amounts to
restricting $a$ to be an $rs$-th 
root of unity. This can be interpreted
geometrically as restricting the positions of the D-branes to the vertices of a 
regular $rs$-gon. In other words, imposing on the boundary also the rational 
symmetries restricts the D-brane moduli to take their values only in a subset 
of `rational' points. This pattern is familiar from the bulk theories where
typically a \cft\ is rational only at isolated points of its moduli space.

In the more complicated case of $d$ bosons compactified on a torus $T^d$
(see subsubsection \ref{scfb}),
the classifying algebra has to be defined on just one `half' of
the lattice $\Gamma(g,B,M)$ \erf{gbm}, either the left moving or the right
moving part (of course, via multiplication by $M$ both are isomorphic).
For a given background $g,B$, this range depends on the
choice of the \ttype\ $M$. As usual, the representations of the respective
classifying algebras will give rise to additional continuous moduli, the
\ctype s.

Note that in some of the previous examples we have successfully
applied the concept of a classifying algebra successfully even to theories for 
which the underlying relation \erf{class} was not derived originally:
we used it also for (generalized) Dirichlet boundary conditions, and we
have applied it also to theories that are not rational. Another extension 
holds for the case of non-trivial torus partition functions $\pi$. 
Using the explicit form of the operator 
product coefficients and the fusing matrices for \sutwo\ \wzwts, 
a classifying algebra for the case where $\om\eq\om_C$
and where the modular invariant $\pi$ is of $D_{{\rm odd}}$-type
has been computed in \cite{prss3,sasT2}. A classifying algebra for 
$\om\eq\om_C$ and general automorphism modular invariants $\pi$
of simple current type was presented
in \cite{fuSc5}. In this case, the structure constants of the classifying 
algebra are still integers, but negative integers occur as well.

\sect{Strings and branes}

This paper is mainly concerned with the structure of (open) \cft. But of
course, once we have established various new features of these theories, we
can also draw conclusions for string theory, for which \cft\ plays the role
of describing consistent vacuum configurations. As an illustration, we 
present in this section a few simple applications of our results, namely 
scattering amplitudes in brane backgrounds, the annulus amplitude, and
comments concerning the properties that must be satisfied by \bc s.

\subsection{Tree level amplitudes in a brane background}
 
To compute scattering amplitudes for a string theory,
the general recipe is to take the correlators of the
underlying \cft, impose the BRST cohomology and then integrate over the
moduli. We briefly illustrate this prescription for the case of
the scattering amplitude for $\no$ open and $\nc$ closed bosonic strings at
`tree' level, i.e.\ on the disk $C$. This amplitude reads
  \be  A(\no,\nc) = c_{\no,\nc}^{} \int\!\!\rmd\mu_x \int\!\!\rmd\mu_z\;
  \LAngle\, \prod_{p=1}^{\no}
  \pHB{A_p}{B_p}{\mu_p}(x_p)\, \prod_{q=1}^{\nc} \phoq\lambda(z_q) \,\RAngle\,.  \ee
Here the various  quantities have the following meaning:
\nxt The prefactor $c_{\no,\nc}^{}$ is a normalization constant.
For general \cfts, the complete calculation of this normalization will be a 
difficult task. Some details about the value of the normalization constant in
the case of free bosons can be found in \cite{fpslr}.
\nxt $\int\! \rmd\mu_x$ is an integral over the positions $x_p \iN\partial C$
     of the open string insertions. It already implements an ordering of these 
     positions; e.g.\ when the world sheet is taken to be the upper half-plane, 
     it is given by
  \be  \rmd\mu_x = \prod_{p=1}^{\no} \rmd x_p\, \theta(x_{p+1}\mi x_p)
  \,,  \ee
while when the world sheet is the unit disk, one has to translate this formula
to the corresponding coordinates on the disk via the map
  \be  z \mapsto \frac{\ii\mi z}{\ii\pl z} \,,  \ee
which sends e.g.\
$0\mapsto1$, $\infty\mapsto-1$, $\pm1\mapsto\pm\ii$.
\nxt $\int\!\rmd\mu_z$ is an integral over the positions $z_q$ of the closed
string insertions, corresponding to the covering surface of the world sheet
and with the M\"obius invariance taken into account properly. Thus
  \be  \rmd\mu_z = ({\rm Vol(M\ddot obius)})^{-1}\, \prod_{q=1}^{\nc} \rmd^2z_q
  \,.  \ee
\nxt The expectation value $\LAngle\cdots\RAngle$ is the {\em correlation
function\/} of the relevant primary fields, i.e.\ of the boundary fields 
$\pHB{A_p}{B_p}{\mu_p}$\,%
\futnote{The boundary labels $A_p$ and $B_p$ are of course not independent,
but satisfy $B_p\eq A_{p+1}$ for $p\eq 1,2\Ldots\no\mi1$ and $B_{\no}\eq A_1$.}
that correspond to the relevant on-shell states of the open strings and the
bulk fields $\pho{\lambda_q}$ that correspond to the relevant on-shell states
of the closed strings. 
Radial ordering of the bulk fields is implicit (and also
an ordering of the boundary fields, but that is already taken care of by the
measure $\rmd\mu_x$).

Using the results of the previous sections, we can express the \cft\
\corfu\ as a linear combination of chiral blocks. The \corfu\ depends, 
of course, on the set $\{(A_p,B_p)\}$ of chosen \bc s. More precisely,
the blocks themselves depend on the chosen \ttype\ $\om$, while
the appropriate linear combination depends on the 
\ctype\ $a\iN\Io$, which is implemented by including the corresponding
prefactors that consist of normalizations and reflection coefficients,
analogously to \erf{bA}. In the following we just
write one of these blocks; the summation over the relevant allowed 
intermediate sectors (including multiplicities, which we suppress as well; 
compare the remarks after \erf{cvo}) must be restored at the end.
In terms of the fields, this means in particular that we have to
express the bulk fields through \cvo s as in \erf{OO}, i.e.\ we write
$\prod_{q=1}^{\nc} \phoq\lambda(z_q) = \prod_{q=1}^{\nc} 
\phd{\lambda_q}(z_q) \odot \prod_{q=1}^{\nc} \phd{\Bar\lambda_q}(I z_q)$.
We may also specify the basis of blocks, e.g.\ the one for which
the arguments are radially ordered. In that case the ordering in the first
product is `opposite' to the one in the second product, analogously as in
the formula \erf{Gamma}. 

To evaluate the amplitude further, one may also use the M\"obius 
transformations to fix one of the insertion points, say $p_1$ to $z_1\eq0$ 
(and hence $I(z_1)$ to $\infty$), so that one is left with the product of 
$2(n_c{-}1)$ \cvo s sandwiched between highest weight vectors 
$|\psi_{\Bar\lambda_1}^{}\rangle$ and $|\psi_{\lambda_1}^{}\rangle$.
Moreover, by imposing the intertwining property
  \be   \Illl(z) = \zeta^{L_0-1}\,\Illl(z/\zeta)\,\zeta^{-L_0} \ee
for Virasoro-primary \cvo s of conformal weight $\Delta\eq1$ one may
scale all other positions by $1/z_{n_c}$. In the case of Neumann \ttype\ and 
$\no\eq 0$ (i.e.\ no boundary insertions), e.g., the \cb\ then reads
  \be  (z_{n_c}\,I(z_{n_c}))_{}^{-({n_c}-2)} \,
  \Langle \psi_{\Bar\lambda_1}^{} {\mid}\,
  \Llb \prod_{q=2}^{n_c} \phd{\Bar\lambda_q}^{}(I(\Frac{z_q}{z_{n_c}}))\Lrb \,
  \Llb (I(z_{n_c}))^{-L_0}_{} (z_{n_c})^{L_0}_{}\Lrb \,
  \Llb \prod_{q=2}^{n_c} \phd{\lambda_q}^{}(\Frac{z_q}{z_{n_c}}) \Lrb
  \,{\mid} \psi_{\lambda_1}^{} \Rangle \,.  \ee
Finally one may separate the three parts of the operator appearing here by
using the state-field correspondence so as to insert twice a summation over a 
complete set of states. This displays nicely the over-all structure of the \cb.
Of course, when one is dealing with a generic \cft, then it is a difficult 
task to write down the \cb, and thereby the string scattering amplitude, more 
explicitly. In contrast, when one specializes to the theory of free bosons, a 
lot of simplifications occur, by which one can reduce the results above to the
corresponding formulas for free bosons which can be found in \cite{fpslr}.

\subsection{The annulus amplitude}\label{scp}

As a particular example, we consider the annulus vacuum-to-vacuum
amplitude with Neumann \ttype\ on both boundary components. This
can be calculated as
  \be  A^{ab} = \int_0^\infty \Frac{\rmd t}{t^2}\, A^{ab}(t)  
  = \int_0^\infty \Frac{\rmd t}{t^2}\, \sum_{\mu,\nu\in\Iset} A^{ab}_\mu
    S_{\mu\nu}^{} \chii_\nu(\Frac\ii{2t})  
  = \Frac12 \int_0^\infty \!\rmd u\, \sum_{\mu,\nu\in\Iset} A^{ab}_\mu
    S_{\mu\nu}^{} \chii_\nu(\ii u) \ee
with $A^{ab}_\mu$ as defined in \erf{aa}.
This amplitude, in contrast to closed string amplitudes, is {\em not\/} finite
by itself, since massless or tachyonic states lead to contributions to
the characters over which the moduli integral diverges. The common strategy
is then the following. After choosing a standard normalization of the moduli 
integrals, one adds up the integrands of the various amplitudes for
all surfaces (both oriented and unoriented) that have the same
Euler number, where one allows in addition for certain multiplicities,
the so-called {\em\cp\/}multiplicities.\,%
\futnote{The \cp multiplicities are responsible for the gauge symmetries
in open string theories, see e.g.\ \cite{pach,masa}.
The reader may be accustomed to having only a single type of \cp label;
this corresponds to the situation of 
uncompactified free bosons (i.e., with diagonal torus partition function). 
In contrast, in the generic case there will be an (essentially \cite{prss}
independent) \cp multiplicity for each \ctype\ $a$, that is, once the \ttype\ 
has been fixed, one for each allowed \bc.} More precisely, essentially one 
counts each boundary condition $a$ not just once, but $N_a$ times, which 
amounts to multiplying the annulus coefficients $A^{ab}_\mu$ by factors of 
$N_aN_b$.
One then requires that after inclusion of these multiplicities the sum of the
integrands is an integrable function on the moduli space. By imposing
this cancellation of the `tadpoles' one determines (at least partly) the
values of the \cp multiplicities. This constitutes one of the few known ways
for determining which sectors have to be included
in a consistent string theory. Let us point out that the determination
of \cp multiplicities is a problem of string theory and cannot
even be formulated in pure \cft\ terms. (Also note that there is a priori 
no reason why a consistent solution should exist at all.)

\subsection{Boundary conditions}

Our next remark concerns the allowed boundary conditions. In the previous
sections, we have always implicitly assumed that the boundary conditions 
preserve the full symmetries in the bulk (possibly in a twisted way), which was
reflected by the fact that we used the full chiral algebra to define the
boundary blocks. This is actually a very strong condition, and for specific 
applications it might be necessary to relax it. Indeed, in the application of 
open \cft\ to \twodim\ critical phenomena there is typically no reason to 
require that the boundary preserves more symmetries than just the Virasoro 
algebra, and in special situations it may even be possible to dispense of the
preservation of the full Virasoro algebra.

In string theory, it is usually argued that the boundary should preserve
the symmetry that is gauged, i.e.\ the Virasoro algebra in the bosonic string,
\resp\ its corresponding super extensions for the various types of
superstrings. It seems to us, however, that this requirement is a bit
too restrictive. Namely, when we work in the covariant description of string 
theory, we must supplement the `internal' \cft\ by
a ghost system and take a BRST cohomology on it. Accordingly, the boundary
blocks have to be complemented by boundary blocks in the ghost theory.
When doing so, boundary states for which the (super-)Virasoro algebra is 
preserved only up to BRST-exact 
terms seem to be perfectly admissible as well. 
(It has been suggested \cite{polc3} that this happens in the presence of 
\nontriv\ Ramond\hy Ramond background charges.) A similar phenomenon arises in
the description of the fixed point sectors in gauged \wzwts. These sectors
exist due to certain selection rules. In view of this fact, it is
tempting to conjecture that a similar algebraic structure underlies the
D-branes that resolve the conifold singularities; recall that at the
conifold point a subset of the `perturbative' states becomes infinitely
massive and decouples, leading to an effective selection rule.

\sect{Outlook}

In this paper we have established the structures in open \cft\ that are
relevant to the description of D-branes for arbitrary \cft\ backgrounds.
We do not repeat any of our results here; rather, we comment on several open 
problems and propose further lines of research. We start with questions on the 
level of chiral \cft. Clearly, the system of D-brane blocks 
deserves further study. 
First, one should try to find a still more explicit characterization of
the fusion rule automorphisms that lead to consistent \ttype s.
Moreover, one should explore in detail to what extent the properties of 
ordinary \cb s are realized for the system of D-brane blocks as well. 
In particular, one
should establish the existence of a \KZ\ connection, and factorization
rules as well as an analogue of the Verlinde formula should be formulated 
and proven. In this context we remark that in factorization constraints
typically blocks of several different \ttype s are involved.
It is also worth stressing that the \KZ\ connection is a central piece of
structure for chiral \cft. {}From the existence of a generalized
\KZ\ connection one could in particular derive the existence of analogues of
braiding and fusing matrices for general \ttype s. The latter will be a
crucial input in the proof of the existence of a
classifying algebras for arbitrary \ttype.

On the level of full \cft, the structure of the set $\Io$ of \ctype s\ that 
are allowed for a given \ttype\ $\om$ remains to be clarified. As was argued 
in \cite{prss2,prss3}, this set
is in particular endowed with a conjugation; this conjugation is not unique, 
which leads to different possibilities for \cp groups in the
corresponding string theories. A more detailed description of the set of 
\ctype s\ would hopefully lead to a natural characterization of the consistent
conjugations. In this context, it is a striking observation \cite{prss3,fuSc5}
that in the case of non-trivial torus partition functions structures in the 
space of boundary conditions show up which closely resemble the ones implied
by modular invariance in closed \cft.
Finally, also the description of D-branes that wrap a cycle more than once
should be made more explicit. Recall that an $n$-fold wrapping of a D-brane
corresponds to a vector bundle of rank $n$ over its world volume. In
particular, as we have seen, the description of those vector bundles which do
not split into
a direct sum of line bundles requires a careful analysis of the embedding
of (subsectors of) chiral \cfts\ into some other chiral \cft.

Let us also comment on a few aspects on the level of string theory.
The \cft\ structures established in this paper provide us with a
description of the usual Dirichlet $p$-branes. Clearly, other
intriguing questions are raised by the study of the other types of branes that
are believed to be present in string theories, like e.g.\ the Neveu\hy Schwarz
five-brane. For some recent attempts to describe open strings in the
background of this brane, we refer to \cite{fogp,bist}.

Finally, it would be important to find a better procedure than tadpole
cancellation to determine \cp multiplicities. Several theories are known
\cite{zwar,prapriv} in which these conditions do not allow for any solution,
but which otherwise seem to be consistent.
This problem is connected to the problem of finding a clearly formulated
principle which tells us which sectors have to be included in a string
theory with which multiplicities. So far, this type of question is answered
mostly by invoking auxiliary arguments which cannot be formulated in a purely
stringy way, like e.g.\ anomaly freeness of some effective field theory.
Such a principle would also be a necessary prerequisite for a 
proof of string dualities beyond the BPS level.

\bigskip
\bigskip
\noindent {\bf Acknowledgement:} \\
We thank P.\ Bantay, M.\ Bianchi, D.\ Cangemi, P.\ di Vecchia, G.\ Felder,
J.\ Fr\"ohlich, O.\ Lech\-tenfeld, G.\ Pradisi, A.K.\ Raina, A.\ Recknagel,
A.\ Sagnotti, V.\ Schomerus, R.\ Silvotti, and Y.\ Stanev
for helpful discussions, and A.\ Recknagel and V.\ Schomerus for useful 
comments on an earlier version of the manuscript.
\vskip3em
\ifnum\draftcontrol=0 \newpage \fi

 \def\wb{\,\linebreak[0]} \def\wB {$\,$\wb}
 \def\Bi{\bibitem }
 \newcommand\Erra[3]  {\,[{\em ibid.}\ {#1} ({#2}) {#3}, {\em Erratum}]}
 \newcommand\BOOK[4]  {{\em #1\/} ({#2}, {#3} {#4})}
 \newcommand\J[5]   {\ {\sl #5}, {#1} {#2} ({#3}) {#4} }
 \newcommand\Prep[2]  {{\sl #2}, preprint {#1}}
 \newcommand\inBO[7]  {\ {\sl #7},
                      in:\ {\em #1}, {#2}\ ({#3}, {#4} {#5}), p.\ {#6}}
 \newcommand\vypF[5]  {\ {\sl #5}, {#1} [FS{#2}] ({#3}) {#4}}
 \newcommand\Habl[2]  {\ {\sl #2}, Habilitation thesis (#1)}
 \def\anop  {Ann.\wb Phys.}
 \def\aspm  {Adv.\wb Stu\-dies\wB in\wB Pure\wB Math.} 
 \def\comp  {Com\-mun.\wb Math.\wb Phys.}
 \def\crap  {C.\wb R.\wb Acad.\wb Sci.\wB Paris (S\'erie I -- Math\'ematique)}
 \def\foph  {Fortschr.\wb Phys.} 
 \def\ijmp  {Int.\wb J.\wb Mod.\wb Phys.\ A}
 \def\jams  {J.\wb Amer.\wb Math.\wb Soc.} 
 \def\joag  {J.\wB Al\-ge\-bra\-ic\wB Geom.}
 \def\jodg  {J.\wb Diff.\wb Geom.} 
 \def\jpaa  {J.\wB Pure\wB Appl.\wb Alg.} 
 \def\lemp  {Lett.\wb Math.\wb Phys.}
 \def\mpla  {Mod.\wb Phys.\wb Lett.\ A} 
 \def\nuci  {Nuovo\wB Cim.}
 \def\npbp  {Nucl.\wb Phys.\ B (Proc.\wb Suppl.)}
 \def\nupb  {Nucl.\wb Phys.\ B}
 \def\npbF  {Nucl.\wb Phys.\ B\vypF}
 \def\phlb  {Phys.\wb Lett.\ B}
 \def\phrd  {Phys.\wb Rev.\ D} 
 \def\phrl  {Phys.\wb Rev.\wb Lett.}
 \def\rvmp  {Rev.\wb Math.\wb Phys.}
 \def\sebo  {S\'emi\-naire\wB Bour\-baki}
 \def\NY    {{New York}} 
 \def\PR    {{Providence}} 
\newcommand\geap[2] {\inBO{Physics and Geometry} {J.E.\ Andersen, H.\
            Pedersen, and A.\ Swann, eds.} \MD\NY{1997} {{#1}}{{#2}} }
\newcommand\kniz[2] {\inBO{The Physics and Mathematics of Strings,
            Memorial Volume for V.G.\ Knizhnik} {L.\ Brink, D.\ Friedan, and
            A.M.\ Polyakov, eds.} \WS\Si{1990} {{#1}}{{#2}} }
\newcommand\nqft[2] {\inBO{
            Nonperturbative Quantum Field Theory} {G.\ 't Hooft, A.\
            Jaffe, G.\ Mack, P.K.\ Mitter, and R.\ Stora, eds.}
            \PL\NY{1988} {{#1}}{{#2}} }
\newcommand\phgt[2] {\inBO{Physics, Geometry, and Topology}
            {H.C.\ Lee, ed.} \PL\NY{1990} {{#1}}{{#2}} }
\def\A      {Algebra}
\def\AMS    {{American Mathematical Society}} 
\def\AP     {{Academic Press}} 
\def\BIR    {{Birk\-h\"au\-ser}}
\def\Bo     {{Boston}}
\def\Ca     {{Cambridge}}
\def\compac {compactification}
\def\con    {conformal }
\def\Con    {Conformal }
\def\CUP    {{Cambridge University Press}}
\def\CY     {Cala\-bi\hy Yau }
\def\dim    {dimension}
\def\eq     {equa\-tion}
\def\furu   {fusion rule}
\def\Intro  {Introduction }
\def\jf     {J.\ Fuchs}
\def\KS     {Kazama\hy Suzuki }
\def\MD     {{Marcel Dekker}}
\def\modinv {modular invarian}
\def\Modinv {Modular invarian}
\def\parfu  {partition function}
\def\PL     {{Plenum Press}}
\def\q      {quantum }
\def\Q      {Quantum }
\def\qzn    {quantization}
\def\Rep    {Representation}
\def\RI     {Riemann}
\def\Si     {{Singapore}}
\def\stt    {string theory}
\def\stts   {string theories}
\def\susy   {supersymmetry}
\def\sym    {symmetry}
\def\syms   {sym\-me\-tries}
\def\trfo   {transformation}
\def\voa    {vertex operator algebra}
\def\vop    {vertex operator}
\def\Vop    {Vertex operator}
\def\WS     {{World Scientific}}
\def\WZW    {Wess\hy Zu\-mino\hy Wit\-ten }

\end{document}